\documentclass[twocolumn]{aastex631}
\usepackage{graphicx}
\usepackage[flushleft]{threeparttable}
\usepackage{tabularx}
\usepackage{booktabs}
\newcolumntype{Y}{>{\centering\arraybackslash}X}
\usepackage{multirow}
\usepackage{amsmath}
\usepackage{times}

\defcitealias{Kennicutt12}{KE12}
\defcitealias{Aravena19}{A19}
\defcitealias{Birkin21}{B21}
\defcitealias{Tacconi20}{T20}

\newcommand\gildas{\textsc{gildas}}

\submitjournal{ApJ}

\shorttitle{CO in Faint DSFGs}
\shortauthors{Nicandro Rosenthal et al.}
\graphicspath{{../figures/}{figures/}{../../plots/}}

\begin{document}

\title{Molecular Gas Detections in Eight Faint DSFGs with Red NIR Colors at $\mathbf{z \sim 1.2-2.5}$}

\correspondingauthor{Michael J. Nicandro Rosenthal}
\email{rosenthal@astro.wisc.edu}

\author[0000-0003-3910-6446]{Michael J. Nicandro Rosenthal}
\affiliation{Department of Astronomy, University of Wisconsin--Madison, 475 N. Charter Street, Madison, WI 53706, USA}

\author[0000-0003-3910-6446]{Stephen J. McKay}
\affiliation{Department of Physics, University of Wisconsin--Madison, 475 N. Charter Street, Madison, WI 53706, USA}

\author[0000-0002-3306-1606]{Amy J. Barger}
\affiliation{Department of Astronomy, University of Wisconsin--Madison, 475 N. Charter Street, Madison, WI 53706, USA}
\affiliation{Department of Physics and Astronomy, University of Hawaii, 2505 Correa Road, Honolulu, HI 96822, USA}
\affiliation{Institute for Astronomy, University of Hawaii, 2680 Woodlawn Drive, Honolulu, HI 96822, USA}

\author[0000-0002-6319-1575]{Lennox L. Cowie}
\affiliation{Institute for Astronomy, University of Hawaii, 2680 Woodlawn Drive, Honolulu, HI 96822, USA}




\begin{abstract}
\noindent
We present a NOEMA survey of CO(3--2), CO(4--3), and [C\,{\sc i}]($^3$P$_1$--$^3$P$_0$) in eight faint (average $S_{\rm 850 \mu m} = 2.3$\,mJy) dusty star-forming galaxies (DSFGs) at $z = 1.2-2.5$. We used a NIR flux-color cut to match faint SCUBA-2 sources to red stellar counterparts with existing spectroscopic redshifts, allowing us to target CO lines at known frequencies. We obtained seven new CO detections and a serendipitous [C\,{\sc i}] detection in an off-axis source, and measured molecular gas masses of $M_{\rm mol} = (6-22)\times10^{10}\,(\alpha_{\rm CO}/3.6)\,{\rm M}_\odot$ from these lines. We performed UV-to-mm SED fits to measure the SFRs and stellar masses of our sample, and compared these with two other $z = 1-3$ CO samples from the literature. The CO detections have constant depletion times of $t_{\rm dep} \sim 500$\,Myr, with no evidence for correlation between $t_{\rm dep}$ and redshift or main-sequence offset. We find that low-mass ($M_\star \lesssim 10^{11}\,{\rm M}_\odot$), starbursting galaxies have gas fractions and depletion times twice as high as predicted by molecular gas scaling relations, which may indicate that $M_{\rm mol}$ is systematically over-estimated in this population, possibly due to decreased $\alpha_{\rm CO}$ or increased CO excitation compared to the well-studied massive and/or main-sequence DSFG population.
\end{abstract}



\section{Introduction} \label{sec:intro}

Roughly half of the star formation across cosmic time is obscured by dust \citep[e.g.,][]{Madau14, Zavala21} and re-radiated at submillimeter/millimeter (submm/mm) wavelengths. While the brightest dusty star-forming galaxies (DSFGs) have the highest individual star formation rates (SFRs) of any galaxies, the majority of dust-obscured star formation since $z \sim 2$ has occurred in relatively faint ($S_{\rm 850 \mu m} \lesssim 2$ mJy) DSFGs, which are far more numerous and account for the majority of the total infrared extragalactic background light \citep[e.g.,][]{Chen13, Hsu16, Hsu24}. Measuring the molecular gas content fueling star formation in faint DSFGs is thus a critical component of understanding the processes governing the Universe's star formation history (SFH) as a whole. 

The molecular gas masses ($M_{\rm mol}$) of star-forming galaxies are best measured by observing the rotational emission lines of CO (e.g., \citealp{Greve05, Solomon05, Bothwell13, Tacconi13, Tacconi18, Aravena19}, hereafter \citetalias{Aravena19}; \citealp{Birkin21}, hereafter \citetalias{Birkin21}), chiefly with millimeter interferometers, such as the Northern Extended Millimeter Array (NOEMA) and the Atacama Large Millimeter/submillimeter Array (ALMA). Such surveys have shown that the luminosities of CO lines ($L'_{\rm CO}$) scale monotonically with integrated infrared luminosity \citep[$L_{\rm IR}$, e.g.,][]{Bothwell13, Liu15} at all redshifts, reflecting the fundamental relationship between SFR (traced by $L_{\rm IR}$) and $M_{\rm mol}$ \citep{Schmidt59, Kennicutt98}. Recent works have compiled thousands of CO detections to fit measured molecular gas scaling relations (see \citealt{Tacconi20}, hereafter \citetalias{Tacconi20}, for a review) finding galaxies to be more gas-rich at higher redshifts and lower stellar masses; they also have shorter depletion times ($t_{\rm dep} \equiv M_{\rm mol} / {\rm SFR}$) the higher they are above the star-forming main sequence (SFMS).

Most of these CO surveys, however, are biased toward the brightest, most highly star-forming DSFGs due to their sample selection, especially at $z \gtrsim 1$, since most CO-detected DSFGs have been selected from single-dish submm/mm imaging, and submm/mm flux is directly related to SFR \citep[e.g.,][]{Barger14,Cowie17}. Single-dish telescopes are fundamentally limited by confusion noise \citep[e.g.,][]{Blain98}, such that almost all single-dish sources have ${\rm SFR} \gtrsim 100\,{\rm M_\odot \,yr^{-1}}$. 

An alternative approach is to obtain deep, wide-bandwidth observations over mosaicked fields, and thus forgo target selection altogether. The most sensitive such survey is the ALMA Spectroscopic Survey in the Hubble Ultra Deep Field (ASPECS; \citealp{Walter16}; \citetalias{Aravena19}), an ALMA large program that obtained uniform 3\,mm sensitivity across a 5 arcmin$^2$ survey area. They detected CO emission in 18 galaxies, which spanned a wider range of stellar masses, SFRs, and SFMS offsets than those in pre-selected CO surveys \citepalias{Aravena19}. While successfully circumventing the biases associated with pre-selected DSFG samples, this approach is extremely inefficient compared to targeted CO surveys. Global inferences about the molecular gas properties of faint DSFGs inferred from surveys such as this are limited both by the small sample size and by potential bias due to cosmic variance, given the small survey area. Constructing large samples of CO in faint DSFGs over larger areas is therefore necessary to characterize properly the CO in the faint DSFG population.

In this work, we present a pilot survey with NOEMA that uses a targeted approach but leverages the deepest available single-dish imaging and pre-JWST optical/infrared (OIR) spectroscopy to restrict our sample to galaxies with low $L_{\rm IR}$. We select the faintest sources from confusion-limited imaging with the Submillimeter Common-User Bolometer Array 2 \citep[SCUBA-2;][]{Holland13} and employ a near-infrared (NIR) color selection to determine accurately the OIR counterparts to the faintest SCUBA-2 sources \citep{Barger23,McKay24}. We use this counterpart selection to also target only DSFGs with known spectroscopic redshifts, greatly improving the efficiency of our survey compared to deep, blind surveys like ASPECS.

We organize the paper as follows: In Section \ref{sec:sample}, we describe our sample selection and observing strategy, including the NIR color selection we use to identify OIR counterparts with existing spectroscopic redshifts. We describe our 2\,mm NOEMA observations, data reduction, and analysis of the line and continuum emission in Section \ref{sec:data}, and the results of this analysis in Section \ref{sec:results}. We derive the physical properties of our sample in Section \ref{sec:physical}, including deriving molecular gas masses from our observed CO and [C\,{\sc i}] lines. 
Finally, in Section \ref{sec:discussion}, we compare 
our CO sample with several larger CO surveys using ALMA and NOEMA and with predictions from molecular gas scaling relations, and we discuss prospects for future surveys of molecular gas that leverage NIR color selection. We summarize the paper in Section \ref{sec:summary}.

All calculations throughout this work assume a concordance flat $\Lambda$CDM cosmology with $H_0 = 70~{\rm km~s^{-1}~Mpc^{-1}}$, $\Omega_{\rm m,0} = 0.3$, and $\Omega_{\rm \Lambda,0} = 0.7$. Masses and SFRs are scaled to a \citet{Chabrier03} initial mass function (IMF) unless otherwise stated. All magnitudes are in the AB system.

\section{Data and Sample Selection}
\label{sec:sample}

\subsection{Confusion-Limited SCUBA-2 Imaging}

To select faint DSFGs, we use the 850\,$\mu$m source catalog from the Submillimeter Perspective on the GOODS Fields (SUPER GOODS) survey of the GOODS-N \citep{Cowie17} as our parent DSFG sample. This imaging covers an area of $\sim$400 arcmin$^2$ centered on the 2\,Ms \textit{Chandra} image \citep{Alexander03} and the original HST GOODS-N field \citep{Giavalisco04}. It includes a circular region of radius $\sim$4$'$ (area $\sim$50 arcmin$^2$) where the instrumental noise reaches the confusion limit, such that the faintest $4\sigma$ detections have fluxes of only $S_{\rm 850 \mu m} = 1.4$\,mJy. These are the faintest DSFGs that can be detected by SCUBA-2 without gravitational lensing \citep[e.g.,][]{Cowie22}.

\subsection{Ancillary OIR Photometry and Spectroscopy}
\label{sec:candels}

For OIR data, we use the multiwavelength photometric catalog of \citet{Barro19}. This catalog uses the HST/Wide Field Camera 3 (WFC3) F160W image from the Cosmic Assembly Near-infrared Deep Extragalactic Legacy Survey \citep[CANDELS;][]{Grogin11, Koekemoer11} for source detection, reaching a 5$\sigma$ limiting magnitude of ${\rm F160W} = 27.3$ across a survey area of 171 arcmin$^2$ centered on the GOODS-N. It includes photometry in 23 broadband filters from UV to far-infrared (FIR) wavelengths and contains 33,961 secure galaxy detections, i.e., objects that were not flagged as having poor data quality and not classified as stars. The CANDELS high-level science products\footnote{\url{https://archive.stsci.edu/hlsp/candels}} also include a spectroscopic catalog, which was compiled from a variety of spectroscopic surveys of the GOODS-N. In this work, we use the \texttt{v2} spectroscopic catalog from \citet{Kodra23}, which contains 3206 spectroscopic redshifts.

\subsection{Red Selection to Identify Counterparts to DSFGs}
\label{sec:counterparts}

Mid-infrared (MIR; \citealp[e.g.,][]{Pope06}) or radio \citep[e.g.,][]{Barger12} sources have been used for counterpart matching to DSFGs in the past. However, since the advent of JWST, focus has turned to using red sources. \citet{Barger23} found that a simple JWST/NIRCam selection of $f_{\rm F444W} > 1\,\mu{\rm Jy}$ and $f_{\rm F444W}/f_{\rm F150W} > 3.5$ identified the ALMA counterparts of most SCUBA-2 sources; \citet{McKay24} later showed this selection to be $\sim$95\% accurate. The unprecedented depth of NIRCam means that the number of DSFGs that can be identified this way far exceeds the number with MIR or radio counterparts, as well as having better spatial resolution and being less susceptible to active galactic nucleus (AGN) contamination than most MIR or radio priors. This makes NIR color selection especially promising for the study of faint, lower-mass DSFGs \citep[e.g.][]{McKay25, Barrufet25}.

In this work, we match the SUPER GOODS sources with their CANDELS counterparts by using a NIR color selection with the Spitzer/IRAC 4.5\,$\mu$m (hereafter IRAC2) and HST/WFC3 F160W filters:
\begin{equation}
    f_{\rm IRAC2} \geq 1\,\mu{\rm Jy} ~\wedge~ \frac{f_{\rm IRAC2}}{f_{\rm F160W}} > 3.5 \,. \label{eqn:red}
\end{equation}
We hereafter refer to this as our \textit{red selection}. This selection was chosen to be analogous to the NIRCam selection of \citet{Barger23}. 

\begin{figure}[t]
    \centering
    \includegraphics[width=\linewidth]{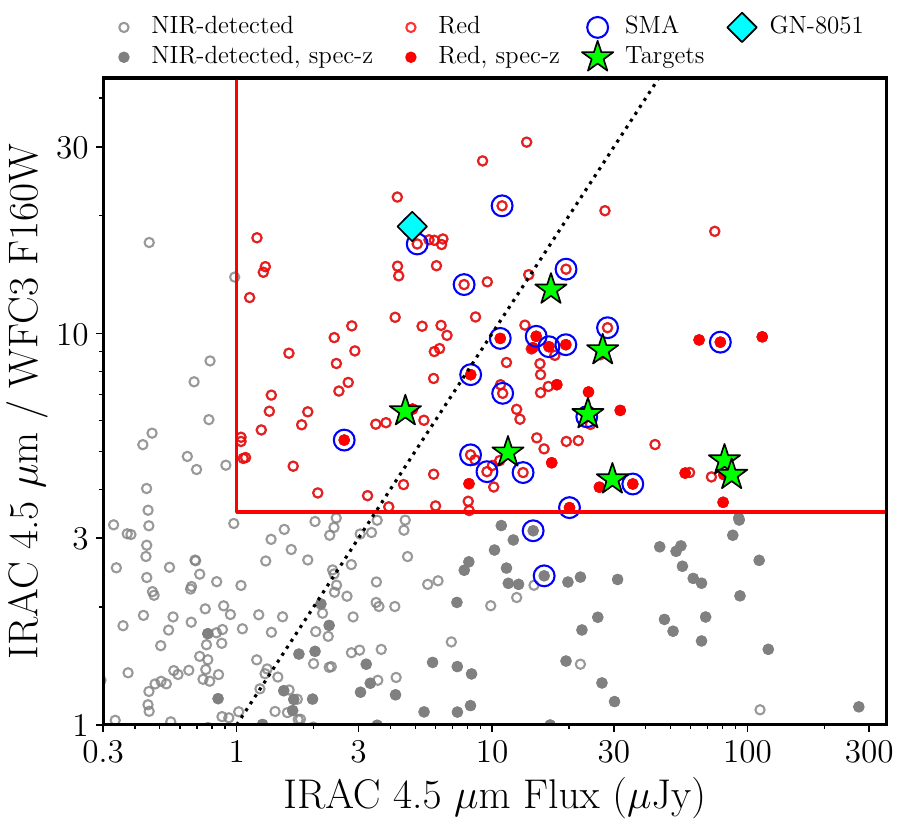}
    \caption{HST/WFC3 F160W and Spitzer/IRAC 4.5\,$\mu$m color-magnitude diagram showing the criterion used for our sample selection. The circles in this figure are \citet{Barro19} CANDELS galaxies within 5$''$ of a \citet{Cowie17} 850\,$\mu$m position. The red solid lines show our red selection (Equation \ref{eqn:red}), with red and gray points showing galaxies that do and do not satisfy this selection, respectively. Filled and open circles are points with and without spectroscopic redshifts, respectively. The large blue circles show CANDELS galaxies wtihin $1''$ of an SMA position from \citet{Cowie17}. Our NOEMA targets are shown as green stars, while the cyan diamond shows the serendipitous detection GN-8051 (see Section \ref{sec:other_sources}). The black dotted line shows an arbitrary cutoff of 1\,$\mu$Jy (23.9 AB mag) in F160W, below which very few sources have spectroscopic redshifts.}
    \label{fig:selection}
\end{figure}

In Figure~\ref{fig:selection}, we plot all of the \citet{Barro19} galaxies that lie within $5''$ of a SUPER GOODS 850 $\mu$m source, with our red selection shown by the red lines. This $5''$ radius is motivated by ALMA follow-up of SCUBA-2 sources in the GOODS-S by \citet{Cowie18}, who found that $>$90\% of SCUBA-2 sources have at least one ALMA counterpart within this radius. We mark the 22 CANDELS galaxies within $1''$ of an 860\,$\mu$m Submillimeter Array (SMA) position from \citet{Cowie17} with open blue circles; 20/22 (91\%) of these satisfy our red selection. 

We note that while the red selection we employ is similar to the ``$H$--IRAC Extremely Red Object'' \citep[``HIERO'';][]{Wang16,Wang19} criterion (${\rm F160W - IRAC2} < 2.25$ mag, or $f_{\rm IRAC2} / f_{\rm F160W} > 7.9$), our inclusion of galaxies with less red colors, that is, $3.5 \leq f_{\rm IRAC2} / f_{\rm F160W} \leq 7.9$, is better at selecting the OIR counterparts of less massive and/or less dusty DSFGs at $z < 3$ \citep{McKay25}. This is demonstrated by the fact that 8/22 of the open blue circles (and 6/8 of the targets in this work; see Section~\ref{sec:final_sample}) are not HIEROs but are selected by our method.

In total, 127 CANDELS galaxies satisfy our red selection and fall within $5''$ of a SCUBA-2 source, making them likely OIR counterparts to the submm sources. These galaxies are shown in Figure~\ref{fig:selection} as red circles, and we select our NOEMA targets from this subsample.

\subsection{NOEMA Targets}
\label{sec:final_sample}

\setlength{\tabcolsep}{9.5pt}
\begin{table*}
	\centering
	\caption{NIR Color-Selected Faint DSFGs}
	\begin{tabularx}{\textwidth}{cc c cccccccc}
		\hline \hline
        \multicolumn{2}{c}{SUPER GOODS} & & \multicolumn{8}{c}{CANDELS} \\
        \cline{1-2} \cline{4-11}
		ID     & $S_{\rm 850\mu m}$ & & ID    & R.A.         & Decl.       & Offset & F160W & $z_{\rm spec}$ & zref & AGN? \\
               & (mJy)              &       &              &     & &  ($''$)      & (AB mag)   \\ 
		(1)    & (2)                & & (3)   & (4)          & (5)         & (6)            & (7)  & (8) & (9) & (10) \\
		\hline
        \multicolumn{10}{l}{\textit{NOEMA Targets}} \\
		183    & $1.4\pm0.3$       & & 13722 & 12:36:57.378 & 62:14:07.99 & 1.35 & 21.8 & 1.4599         & a    & No \\ 
		149    & $2.1\pm0.5$       & & 10601 & 12:37:41.381 & 62:12:51.22 & 2.18 & 20.7 & 1.6017         & a    & Yes \\
        131    & $2.3\pm0.4$       & &  3987 & 12:36:19.115 & 62:10:04.32 & 2.82 & 23.6  & $2.21\pm0.03$  & f   & No  \\
		  110    & $2.9\pm0.3$       & & 10283 & 12:36:34.523 & 62:12:40.99 & 1.15 & 20.8 & 1.224          & c  & Yes \\
         90    & $3.5\pm0.4$       & &  3835 & 12:36:31.281 & 62:09:58.00 & 1.02 & 23.0 & 2.2996         & a,b  & Yes \\
		164    & $1.8\pm0.4$       & & 23490 & 12:36:49.094 & 62:18:14.00 & 1.78 & 22.4 & 2.3197         & a    & No \\
        175    & $1.6\pm0.4$       & &  7630 & 12:37:13.195 & 62:11:45.55 & 2.05 & 24.3 & 2.4139         & a    & No \\
		128    & $2.4\pm0.4$       & & 19876 & 12:36:22.653 & 62:16:29.78 & 3.26 & 22.7 & 2.466          & d,e  & Yes \\ 
        \hline
        \multicolumn{10}{l}{\textit{Other Faint DSFGs (not observed)}} \\
        143 & $2.1\pm0.5$ & &  4295* & 12:36:11.507 & 62:10:33.72 & 1.12 & 22.6 &  2.245 & b & No \\
        162 & $1.8\pm0.5$ & &  7912$^{\#}$ & 12:36:08.824 & 62:11:43.78 & 2.77 & 20.7 &  1.336 & g & No \\
        152 & $2.0\pm0.5$ & &  9568 & 12:36:49.875 & 62:12:33.52 & 4.69 & 22.9 &  2.487 & b & No \\
        118 & $2.7\pm0.4$ & & 17815$^\dagger$ & 12:37:13.668 & 62:15:45.29 & 2.85 & 23.4 & 2.3008 & a & No \\
        \hline
        \multicolumn{10}{l}{\textit{Serendipitously Detected DSFG}} \\
        102    & $3.2\pm0.4$       & & 8051 & 12:37:14.051 & 62:11:56.78 & 0.78 & 25.4 & 2.593 & h & No \\
		\hline
	\end{tabularx}
	\begin{tablenotes}
	\item\textsc{Columns} (1)--(2) are from \citet{Cowie17}'s Table 4, while (3)--(6) are from \citet{Barro19}'s Table 4. The columns are as follows: (1) SCUBA-2 850\,$\mu$m source number; (2) 850\,$\mu$m flux density; (3) CANDELS ID number; (4)--(5) CANDELS position from HST imaging; (6) offset between the SCUBA-2 and CANDELS positions; (7) HST/WFC3 F160W magnitude; (8) spectroscopic redshift; (9) reference for redshift, where a = \citet{Kriek15}, b = \citet{Wirth15}, c = our Keck/DEIMOS [O\,{\sc ii}]\,$\lambda\lambda$\,3727,3729 redshift, d = \citet{Chapman05}, e = \citet{Swinbank04}, f = \citet{Pope08}, g = \citet{Barger08}, and h = \citet{Eisenstein23}; (10) whether the galaxy is associated with a Chandra X-ray source in \citet{Xue16}'s point source catalog (i.e., all galaxies marked ``Yes'' are associated with a moderate luminosity AGN).
    \item*Observed by the PdBI High-$z$ Blue Sequence Survey-2 \citep[PHIBSS-2;][]{Genzel15}.
    \item$^\#$Protected source under the ongoing NOEMA3D large program.
    \item$^\dagger$Observed under NOEMA Project S21CV (P.I. L. Bing; \citealp{Berta25}).
	\end{tablenotes}
    \label{tab:sample}
\end{table*}
\setlength{\tabcolsep}{6pt}

These 127 red-selected CANDELS galaxies are matched to 90 unique SCUBA-2 sources. This average multiplicity of 1.4 is consistent with the number of ALMA counterparts per SCUBA-2 source found by \citet{Cowie18} in the GOODS-S, and suggests of the SCUBA-2 sources may be blends of multiple DSFGs. We limit our sample to only those sources with 1-to-1 matches between red-selected CANDELS galaxies and SCUBA-2 sources. We also require a spectroscopic redshift from \citet{Kodra23} of $1.2 \leq z \leq 3.0$, which is similar to the range probed by \citetalias{Aravena19}. Finally, we limit our sample to faint sources with $S_{\rm 850 \mu m} < 4$\,mJy. 

There are twelve faint DSFGs that satsify all of these criteria. In Table~\ref{tab:sample}, we list the SUPER GOODS and CANDELS ID numbers of these faint DSFGs, their 850 $\mu$m fluxes, their CANDELS positions, the offsets between the SUPER GOODS and CANDELS positions, the F160W magnitudes, and the spectroscopic redshifts with references. We note that for GN-10283, we assume a spectroscopic redshift of $z = 1.224$ based on the detection of the [O\,{\sc ii}]\,$\lambda\lambda$3727,3729 doublet in its Keck/DEIMOS spectrum, rather than the $z = 1.219$ value given by \citet{Kodra23} from earlier, less accurate Keck/LRIS spectra \citep{Chapman05}. Four of the faint DSFGs are also associated with Chandra X-ray point sources, which are consistent with being AGNs with 2--7\,keV luminosities $L_{\rm X} <5\times10^{43}\,{\rm erg\,s^{-1}}$ \citep{Xue16}.

\begin{figure*}[t]
	\centering
	\includegraphics[width=\linewidth]{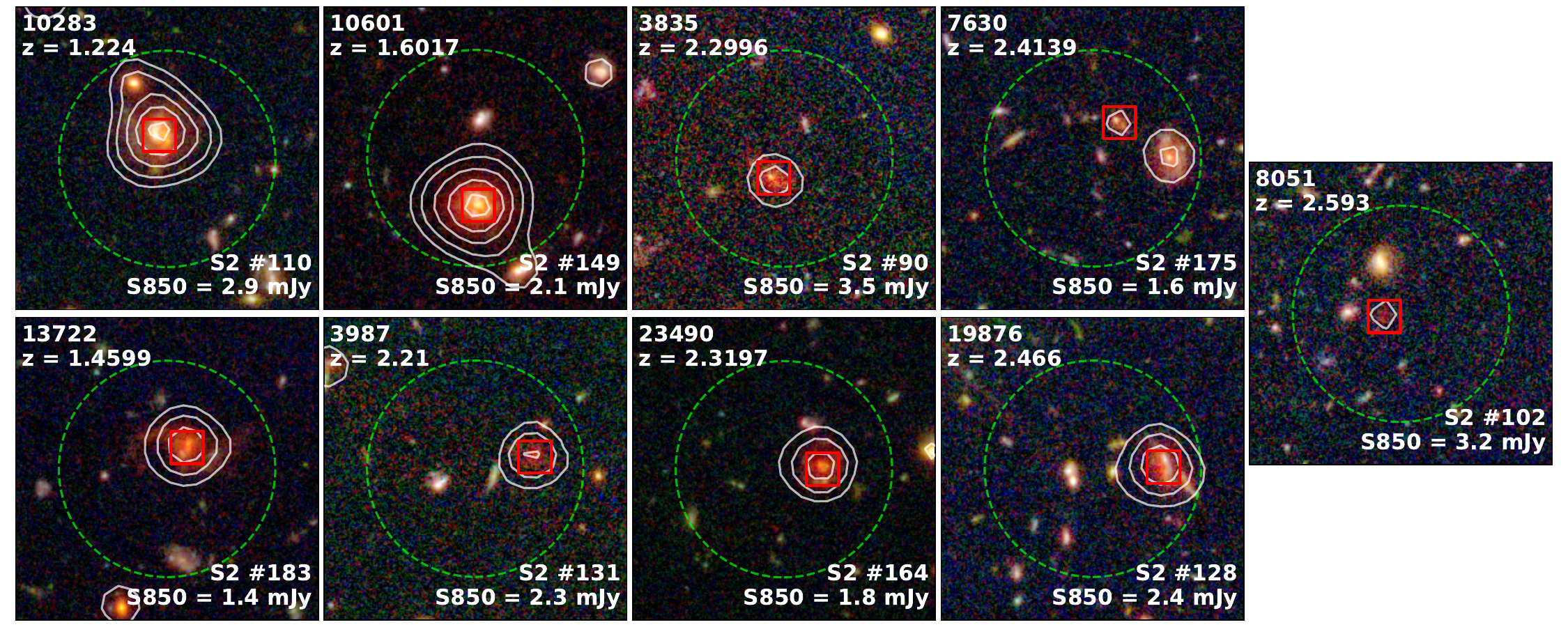}
	\caption{HST three-color thumbnails of our 8 original targets and the serendipitous NOEMA detection GN-8051 (rightmost panel; see Section \ref{sec:other_sources}). All cutouts use the HST filters $rgb$ = F160W/F814W/F435W, with a contour of the Spitzer/IRAC 4.5 $\mu$m emission overlaid in white. Each cutout is $15''$ on a side, centered on the SCUBA-2 position, and the green dashed circle of radius $5''$ is the region within which we search for OIR counterparts. The red square in each thumbnail shows the sole red-selected (Equation \ref{eqn:red}) CANDELS galaxy within this region. The SCUBA-2 ID and 850 $\mu$m flux \citep{Cowie17} are shown in the bottom right corner of each thumbnail, while the CANDELS ID number \citep{Barro19} and spectroscopic redshift \citep{Kodra23} are shown in the top left corner.}
	\label{fig:cutouts}
\end{figure*}

Four of the twelve faint DSFGs were not included in our final NOEMA sample: three are part of past or ongoing NOEMA programs (see the notes for Table \ref{tab:sample}), and the last, GN-9568, would have been the most expensive to observe in our sample due to its high redshift and low $S_{\rm 850 \mu m}$. This left us with a final sample of eight NOEMA targets with $z = 1.224-2.466$, which we show in Figure \ref{fig:selection} as the filled green stars. We show three-color cutouts of our eight NOEMA targets from the CANDELS HST imaging in Figure \ref{fig:cutouts}. Each cutout is centered on the 850\,$\mu$m position and shows our 5$''$ radius for counterpart matching as a green circle. We also plot contours showing the IRAC2 emission in the same region. We mark the red-selected CANDELS galaxies with a red square, showing that each SCUBA-2 source has exactly one such counterpart. 

\subsection{Observing Strategy}

We estimated the expected fluxes of CO transitions in NOEMA bands 1 (3 mm) or 2 (2 mm) at our targets' redshifts by first generating a \citet{Casey12} modified blackbody + MIR power law spectral energy distribution (SED) for each galaxy with fiducial values of $\alpha = 2.0$, $\beta = 1.8$, and $T = 40$ K, normalized to the 850 $\mu$m flux. We integrated these SEDs to obtain FIR luminosities, $L_{\rm FIR}^{\rm 40-400\mu m}$, and used the empirical $L'_{\rm CO(4-3)}$--$L_{\rm FIR}$ relations from \citet{Liu15} to estimate the CO(4--3) fluxes. We then scaled the CO(4--3) values to CO(3--2) and CO(2--1) using the median CO spectral line energy distribution (SLED) from \citetalias{Birkin21}. We found that despite the higher noise in the 2 mm band, the brighter CO fluxes of higher-$J$ emission lines gave us comparable observing time requirements for either 2 mm or 3 mm observations. We observed our targets at 2 mm to maximize the likelihood of obtaining continuum detections, which are helpful for constraining the dust emissivities of these faint DSFGs (see Section \ref{sec:dust_sed}). We therefore targeted two different CO transitions, depending on the redshift of the target galaxy:

\begin{itemize}
	\item CO(3--2) ($\nu_{\rm rest} = 345.786$ GHz) at $z < 2$, and 
	\item CO(4--3) ($\nu_{\rm rest} = 461.041$ GHz) at $z > 2$.
\end{itemize}
In two of the $z > 2$ galaxies, GN-3987 and GN-23490, we targeted both the CO(4--3) transition in the lower sideband (LSB) and the [C\,{\sc i}]($^3$P$_1$--$^3$P$_0$) transition ($\nu_{\rm rest} = 492.161$ GHz; hereafter ``[C\,{\sc i}](1--0)'') in the upper sideband (USB). This gives all of our targeted lines expected sky frequencies of $\nu_{\rm exp} = \nu_{\rm rest,CO} / (1 + z_{\rm spec}) = 133.00-155.83$ GHz. We list the expected frequency for each line in Table \ref{tab:observations}.

Two of our targets had previously been observed in CO surveys using the Plateau de Bure Interferometer (PdBI) and are reported in the compilation of \citet{Bothwell13}. These are GN-3987 (SMM 123618+621007), which has a ``candidate''---rather than secure---CO(3--2) detection, due to its relatively weak ($4.0\sigma$) signal at $z_{\rm CO} = 2.203$ and spatial offset from the SCUBA 850 $\mu$m position \citep{Bothwell13}. In this work, we target its CO(4--3) line to confirm its redshift. Second, GN-10283 (SMM 123634+621241) is a $z = 1.22$ major merger with CO(2--1) \citep{Frayer08}, CO(4--3) \citep{Bothwell13}, and CO(6--5) \citep{Engel10} detections, but not CO(3--2), which we observe in this work. Integration on GN-10283 constituted $\approx$4\% of our usable on-source time.

Our NOEMA observations were conducted between December 2023 and November 2024 in the compact C and D configurations, under NOEMA Projects W23CH (P.I. M. Nicandro Rosenthal) and S24BR (P.I. M. Nicandro Rosenthal). Our observations consisted of nine tracks, totaling 29.5~hr of observing time, with our targets distributed among five observing setups with local oscillator (LO) frequencies $\nu_{\rm LO} = 139-149$ GHz. In Column~(2) of Table~\ref{tab:observations}, we list the setup that we used for each source; sources in the same setup were observed in track-sharing mode to minimize calibration overheads.

\section{Reduction and Analysis}
\label{sec:data}

\setlength{\tabcolsep}{11.5pt}
\begin{table*}
    \centering
    \caption{NOEMA Spectral Line Observations}
    \begin{tabularx}{\textwidth}{rccccccccc}
        \hline \hline
        ID & Setup & Config. & Line & $\nu_{\rm exp}$ & SB & Beam & $t_{\rm int}$ & ${\rm rms}_{\rm ch}$ \\
         & & & & (GHz) & & ($''\times''$, deg.) & (hr) & (mJy\,beam$^{-1}$) \\
        (1) & (2) & (3) & (4) & (5) & (6) & (7) & (8) & (9) \\
        \hline
         \multirow{2}{*}{3987} & \multirow{2}{*}{w23ch001} & \multirow{2}{*}{12C} & CO(4--3)           & 143.63 & L & $1.55\times1.26$, -12.2 & 2.1 & 0.41 \\
              &          & & [C\,{\sc i}](1--0) & 153.32 & U & $1.40 \times 1.13$, -13.9 & 2.1 & 0.51 \\
        13722 & w23ch001 & 12C & CO(3--2)           & 140.57 & L & $1.55\times 1.25$, -9.3 & 2.1 & 0.40 \\
        10601 & w23ch002 & 12C & CO(3--2)           & 133.00 & L & $1.79\times1.10$, 13.9 & 2.2 & 0.43 \\
        \hline
         3835 & s24br001 & 10D & CO(4--3)           & 139.63 & L & $4.01\times2.58$, -55.1 & 1.2 & 0.91 \\
        10283 & s24br001 & 10D & CO(3--2)           & 155.83 & U & $3.61\times2.33$, -54.9 & 0.5 & 1.73 \\
         7630 & s24br002 & 10D & CO(4--3)           & 135.04 & L &  $3.60\times2.43$, -59.4 & 4.2 & 0.34 \\
        8051* & s24br002 & 10D & [C\,{\sc i}](1--0) & 136.98 & L & $3.60\times2.43$, -59.4 & 4.2 & 0.53* \\
        19876 & s24br003 & C/D** & CO(4--3)           & 133.02 & L & $2.83\times 2.24$, 63.0 & 0.6 & 0.88 \\
        \multirow{2}{*}{23490} & \multirow{2}{*}{s24br003} & \multirow{2}{*}{C/D**} & CO(4--3)           & 138.88 & L & $2.82\times2.25$, 64.2 & 0.7 & 0.81 \\
        & & & [C\,{\sc i}](1--0) & 148.25 & U & $2.49\times2.06$, 70.7 & 0.7 & 0.98 \\
        \hline
    \end{tabularx}
    \begin{tablenotes}
    	\item*GN-8051 was a serendipitous detection in the same pointing as GN-7630, located 12.315$''$ from the pointing center. The quoted value of $\sigma_{\rm 20 MHz}$ includes a mean primary beam correction factor across the LSB of this tuning of 1.54.
        \item**Setup s24br003 includes one track taken on 05-Oct-2024 in the 10D configuration, and one track taken on 04-nov-2024 in the 12C--N020+N017 configuration.
    \end{tablenotes}
    \label{tab:observations}
\end{table*}
\setlength{\tabcolsep}{6pt}

We reduced all of our NOEMA data using the standard calibration pipeline in the Continuum and Line Interferometry Calibration (CLIC) component of \gildas, the IRAM data reduction and analysis software suite. The calibrators used varied by track among the following sources: 1125+596, 1214+588, and J1302+690 were used for gain calibration; 0923+392, 3C84, and 3C454.3 for bandpass calibration; and MWC 349 and LKH$\alpha$ 101 for absolute flux calibration. We performed additional manual flagging of both usable tracks in setup s24br003, which had degrading weather conditions at the end of each track that prevented good amplitude and phase versus time fits. For all other tracks, we used the default pipeline output on all of the available data.

Following calibration, we used the CLIC data quality assessment (DQA) tool to flag scans within the calibrated data that have high phase rms noise or large amplitude losses, using the default detection experiment settings. Our final calibrated {\em uv} data tables were generated from the remaining data. In Column~(8) of Table \ref{tab:observations}, we give the total usable integration time on each source after running the DQA. These range between 0.5\,hr for the brightest source (GN-10283) and 4.2\,hr for the faintest source (GN-7630). In total, our survey obtained 13.6\,hr of usable on-source integration time.

\subsection{Imaging}
\label{sec:imaging}

We generated first-look dirty image cubes of each sideband with a targeted emission line in \textsc{mapping}, the imaging and deconvolution suite of \gildas. We first resampled the spectral axis of each data cube by a factor of 10, which yielded channel widths of 20\,MHz ($\Delta v \approx 40\,{\rm km\,s^{-1}}$ at $\nu_{\rm obs} \approx 150$\,GHz). We removed the first and last 10 channels and the central 2 channels, which are susceptible to noise spikes, from the resampled cubes, and imaged these with natural weighting and fixed pixel sizes of $0.25'' \times 0.25''$. Each image cube was generated with $256\times256$ pixels, or $64''\times64''$, ensuring that the imaged region extends up to at least the FWHM of the NOEMA primary beam ($\theta_{\rm HPBW} = 32''$ at 150\,GHz).

The naturally weighted images have spatial resolutions of roughly $1.5'' \times 1.2''$ for observations in the 12C configuration and $3.6'' \times 2.4''$ for observations in the 10D configuration. In Table~\ref{tab:observations}, we give the specific values for each sideband with a targeted emission line. These are sufficiently coarse that we expect all of our targets to be unresolved, given the typical size of dust and gas emission from DSFGs \citep[e.g.,][]{Hodge16}.

For all images and/or image cubes, we measured the rms in the image plane by masking out a circular region centered on the CANDELS position, with a radius equal to three times the synthesized beam FWHM, and then taking the rms of the flux in 1000 randomly drawn pixels in a circular region extending one primary beam FWHM from the phase center. We refer to these noise values measured in the image plane as rms$_{\rm 2D}$ throughout the remainder of this work. In cases where an off-axis source was detected in the image (see Section \ref{sec:other_sources}), we also masked out a circular region of the same size around that source's position.

\subsection{1D Spectra}
\label{sec:spec}

To generate a 1D spectrum for each line in Table \ref{tab:observations}, we started by summing all channels within $\pm$2000 km s$^{-1}$ of the expected frequency of the line, and we identified the brightest spaxel in this image within one synthesized beam FWHM of the CANDELS position. We then extracted a 1D spectrum from this spaxel. We measured the spectral noise on a channel-by-channel basis using the procedure described in Section \ref{sec:imaging}. 

To correct for the lower primary beam response at positions offset from the phase center of each NOEMA pointing, we approximated the response of the NOEMA primary beam as an Airy function. We performed all of our flux and noise measurements on the uncorrected images or visibilities, and we multiplied the rms and flux values by our primary beam correction afterward, in order to preserve signal-to-noise (S/N).

\subsection{Emission Line Fitting}
\label{sec:line_analysis}

\subsubsection{1D Fits}
\label{sec:1d}

To identify emission lines in our data, we performed a two-step Gaussian fitting procedure on the primary beam-corrected 1D spectrum for each line in Table \ref{tab:observations}, using a simple least-squares fitter. The first Gaussian fit was used to approximate the mean frequency, $\bar\nu$, and Gaussian width, $\sigma$, of the emission line. We then fit a first-order baseline to the channels at least $\pm2\sigma$ from the mean of this Gaussian and subtracted this baseline. Finally, we re-fit the baseline-subtracted spectrum to obtain final values of $\bar \nu$, $\sigma$, and $S_{\rm peak}$. 

We measured uncertainties in these parameters by conducting Monte Carlo (MC) simulations of the above fitting procedure with noise-injected synthetic spectra. For each real 1D spectrum, we represented each channel as a Gaussian distribution with a mean equal to the flux in that channel and standard deviation equal to the rms in that channel. We then drew 10\,000 random values from these distributions, resulting in 10\,000 synthetic spectra, which we fit using the same two-step procedure above.

Following \citet{Bothwell13} and \citetalias{Birkin21}, we consider channels with frequencies between $\bar \nu \pm 2 \sigma$ to be those containing line emission. We estimated the 1D S/N of the fit lines to be
\begin{equation}
	{\rm S/N}_{\rm 1D} = \sqrt{N_{\rm ch}} \cdot \frac{\langle S \rangle}{\langle {\rm rms}_{\rm ch} \rangle} \,, \label{eqn:snr1d}
\end{equation}
where $\langle S \rangle$ and $\langle {\rm rms}_{\rm ch}\rangle$ are the average flux and rms, respectively, across all line channels. We classified the fit emission lines as secure detections if ${\rm S/N}_{\rm 1D} \geq 5$, and as tentative detections if $3 \leq {\rm S/N}_{\rm 1D} < 5$.

\subsubsection{Line-Only Fluxes and Images}
\label{sec:line2d}

Following our 1D fits, we fit a 1st-order baseline to the line-free channels in the calibrated visibilities for each detected line, subtracted this baseline from the $uv$ data, and then averaged these continuum-subtracted data to produce a single-channel line-only $uv$ dataset. We imaged these data using the settings described in Section \ref{sec:imaging}, and we measured the uncorrected rms noise in these images following the procedure in Section \ref{sec:imaging}. 

We measured the positions and fluxes of the secure and tentative line detections by directly fitting the line-only $uv$ data using the \texttt{uv\_fit} task, with initial guesses for the fit allowed to vary by $\pm 2''$ from the CANDELS position. We then converted these into integrated flux units by multiplying in the bandwidth used to generate the line-only maps:
\begin{equation}
    I_{\rm line} = \langle S_{\rm line} \rangle_{\rm 2D} \cdot 4\sigma,
\end{equation}
where $\langle S_{\rm line} \rangle_{\rm 2D}$ is the best-fit flux density from \texttt{uv\_fit}, and $\sigma$ is the Gaussian width from our 1D fits. 

For lines detected with ${\rm S/N}_{\rm 2D} > 5$, we deconvolved the images using a \citet{Hogbom74} CLEAN with a gain of 0.1, limited to an elliptical region twice the size of the synthesized beam FWHM centered on the line position from \texttt{uv\_fit}. We stopped each CLEAN when the maximum value in the residual image in this region reached $2\times{\rm rms}_{\rm 2D}$. We recalculated the rms noise on the cleaned images, and we calculated the S/N of the lines in the image plane as
\begin{equation}
    {\rm S/N}_{\rm 2D} = \frac{\langle S_{\rm line} \rangle_{\rm 2D}}{{\rm rms}_{\rm 2D}} \,.
\end{equation}
We describe the results of this analysis in Section \ref{sec:lines}.

\section{Results}
\label{sec:results}

\subsection{Serendipitous DSFG Detections}
\label{sec:other_sources}

\begin{figure*}[tbh]
    \centering
    \includegraphics[width=0.7\linewidth]{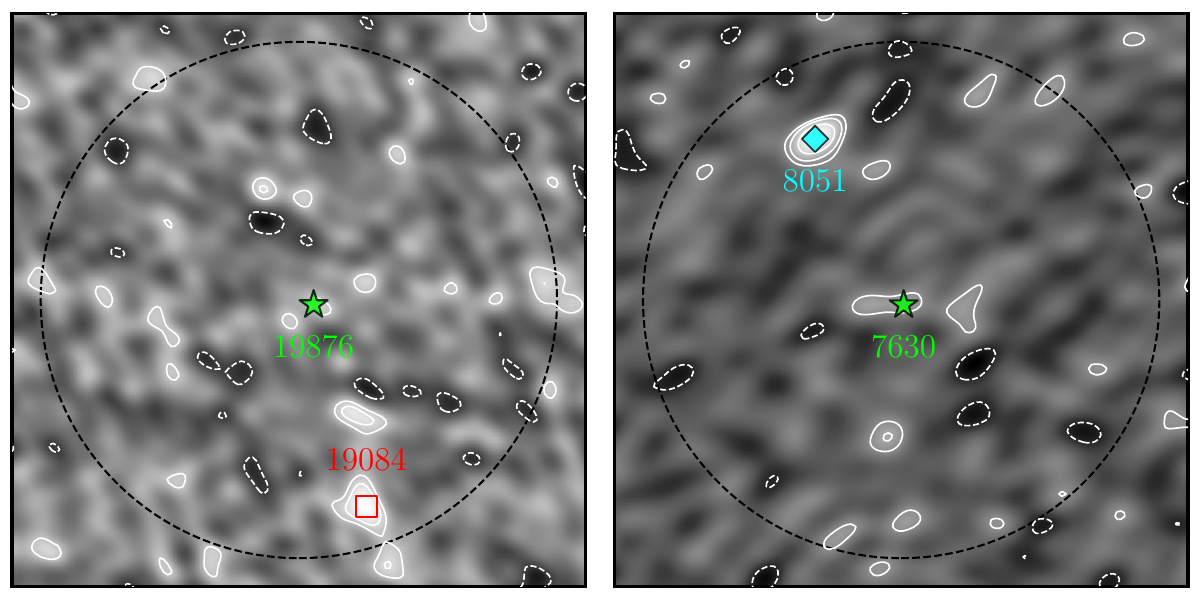}
	\caption{NOEMA 2\,mm continuum images with serendipitous off-axis detections. In both cases, the main target that the pointing is centered on is marked by the green star, as in Figure \ref{fig:selection}, and the black dashed circle shows the extent of the primary beam. Contours are at ${\rm S/N} = -2$ (dashed), 2, 3 and 5. \textit{Left:} Continuum image of the USB, centered on the CANDELS position of GN-19876, with the off-axis source GN-19084 marked by a red square. \textit{Right:} Same, but for the image centered on GN-7630, with GN-8051 marked by the cyan diamond, as in Figure \ref{fig:selection}.}
	\label{fig:dual}
\end{figure*}

Visual inspection of our NOEMA images revealed additional sources within the NOEMA primary beam in two of our pointings (see Figure~\ref{fig:dual}). Each of these additional sources is coincident with an SMA position from \citet{Cowie17}, indicating they are real DSFGs. The two galaxies are

\paragraph{GN-8051} This galaxy was detected at R.A. = 12:37:14, Dec. = +62:11:56.8, offset by $12.7''$ from the pointing center targeting GN-7630. This position is coincident with both the CANDELS galaxy GN-8051 \citep{Barro19} and the SMA position of 850\,$\mu$m source \#102 ($S_{\rm 850\,\mu m} = 3.2\pm0.38$\,mJy) in \citet{Cowie17}. The CANDELS counterpart satisfies our red selection (Equation \ref{eqn:red}; see cyan diamond in Figure~\ref{fig:selection}). GN-8051 was also observed with NIRSpec as part of the JWST Advanced Deep Extragalactic Survey \citep[JADES;][ID = 77760]{Eisenstein23} and found to have $z_{\rm spec} = 2.593$. There are no CO lines at this redshift in our spectral coverage of GN-8051, but the LSB does cover the [C\,{\sc i}](1--0) line at $\nu_{\rm exp} = 136.98$\,GHz. Because it satisfies our selection criteria, we fit the [C\,{\sc i}] and 2\,mm continuum emission of GN-8051 using the procedure outlined in Section \ref{sec:line_analysis}. We list the source's properties at the bottom of Table \ref{tab:sample}, and we include its thumbnail as the rightmost panel of Figure~\ref{fig:cutouts}.

\paragraph{GN-19084} We identified another source at R.A. = 12:36:22.1, Dec. = +62:16:16, offset by $14.6''$ from the pointing center targeting GN-19876. This position is consistent with the CANDELS galaxy GN-19084 \citep{Barro19} and with the SMA position of source \#42 ($S_{\rm 850 \mu m} = 5.4\pm0.41$) in \citet{Cowie17}. It does not have a spectroscopic redshift from \citet{Barro19} or in JADES, and we do not identify any emission lines in this source. We measure and report its 2 mm continuum flux in Section \ref{sec:continuum}.

\subsection{Emission Line Detections}
\label{sec:lines}

\setlength{\tabcolsep}{9pt}
\begin{table*}
	\centering
	\caption{Emission Lines}
    \footnotesize
	\begin{tabularx}{\linewidth}{ccccccccc}
		\hline \hline
		ID    & Line              & S/N$_{\rm 1D}$ & $z$             & FWHM                & $I$                & $L'$ & $L'_{\rm CO(1-0)}$ & $M_{\rm mol}$ \\
		       &                   &                &             & (km\,s$^{-1}$)      & (Jy\,km\,s$^{-1}$) & ($10^{10}$\,K\,km\,s$^{-1}$\,pc$^2$) & ($10^{10}\,{\rm K\,km\,s^{-1}\,pc^2}$) & ($10^{10}\,M_\odot$) \\
		(1) & (2) & (3) & (4) & (5) & (6) & (7) & (8) & (9) \\
		\hline
		10283 & CO(3--2)           &  9.6 & 1.224       & $581^{+60}_{-76}$   & $3.29\pm0.30$ & $2.90\pm0.26$ & $4.60\pm0.97$ & $22.5\pm4.7$ \\[4pt]
		13722 & CO(3--2)           &  9.9 & 1.460  & $369^{+34}_{-37}$   & $0.64\pm0.05$ & $0.79\pm0.06$ & $1.25\pm0.26$ & $6.1\pm1.3$ \\[4pt]
		10601 & CO(3--2)           &  8.5 & 1.599        & $589^{+66}_{-65}$   & $0.71\pm0.07$ & $1.04\pm0.11$ & $1.65\pm0.36$ & $8.1\pm1.8$ \\[4pt]
        \multirow{2}{*}{3987} & CO(4--3)           & 11.0 & 2.215        & $374\pm35$          & $0.72\pm0.06$ & $1.07\pm0.08$ & $3.14\pm0.44$ & $15.4\pm2.2$ \\
          & [C\,{\sc i}](1--0) &  5.5 & 2.215  & $306^{+51}_{-73}$   & $0.39\pm0.05$ & $0.51\pm0.07$ & $2.30\pm0.33$ & $11.3\pm1.6$ \\[4pt]
         3835 & CO(4--3)           & 10.7 & 2.304        & $963^{+73}_{-78}$   & $3.13\pm0.48$ & $4.94\pm0.76$ & $14.54\pm2.81$ & $71.2\pm13.8$ \\[4pt]
		\multirow{2}{*}{23490} & CO(4--3)           &  6.0 & 2.319  & $263^{+52}_{-74}$   & $0.75\pm0.09$ & $1.19\pm0.14$ & $3.50\pm0.57$ & $17.1\pm2.8$ \\
		 & [C\,{\sc i}](1--0) &  1.3* & 2.319*       & 263*        & $<0.24$       & $<0.36$ & $<1.62$ & $<7.9$ \\[4pt]
         7630 & CO(4--3)           &  5.1 & 2.415 & $444^{+364}_{-163}$ & $0.35\pm0.05$ & $0.59\pm0.09$ & $1.74\pm0.34$ & $8.5\pm1.7$ \\[4pt]
		19876 & CO(4--3)           &  1.1* & 2.466*     & 409*        & $<0.54$       & $<0.96$ & $<3.20$ & $<15.7$ \\[4pt]
         8051 & [C\,{\sc i}](1--0) &  3.2 & 2.595 & $510^{+120}_{-212}$ & $0.35\pm0.08$ & $0.60\pm0.14$ & $2.70\pm0.64$ & $13.2\pm3.1$ \\
		\hline
	\end{tabularx}
    	\begin{tablenotes}
            \item{\sc Columns:} (1) CANDELS ID \citep{Barro19}; (2) observed transition; (3) signal-to-noise of the line in the single-spaxel spectrum, given by Equation~\ref{eqn:snr1d}; (4) best-fit redshift; (5) best-fit FWHM with uncertainties from our MC simulations (see Section~\ref{sec:1d}); (6) integrated line flux; (7) line luminosity; (8) luminosity of the fundamental CO(1--0) transition, converted using the average CO SLED from \citetalias{Birkin21} for the CO lines and using the average [C\,{\sc i}](1--0) to CO(1--0) ratio from \citet{Jiao17} for the [C\,{\sc i}] lines (see Section \ref{sec:gas} for details); (9) molecular gas mass, assuming $\alpha_{\rm CO} = 3.6$.
			\item*---For upper limits, Column (3) gives the 1D signal-to-noise of the best-fit Gaussian, but Columns (4) and (5) give the \textit{fixed} redshift and FWHM used to generate the ``line'' images on which the upper limits in Columns (7)--(9) are measured.
		\end{tablenotes}
	\label{tab:lines}
\end{table*}
\setlength{\tabcolsep}{6pt}

\begin{figure*}[htb]
	\includegraphics[width=\linewidth]{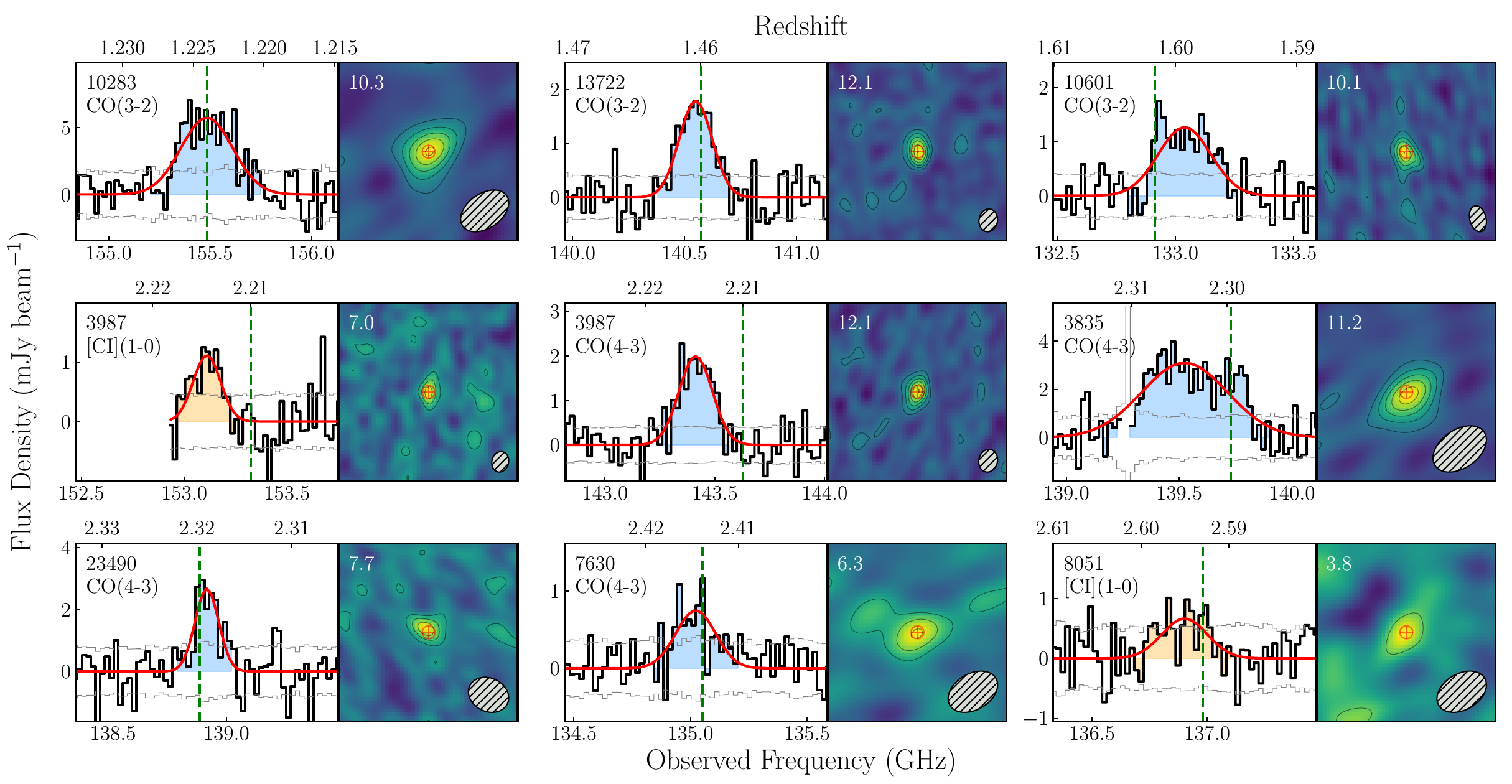}
	\caption{NOEMA spectra and images of the nine detected emission lines in our sample. In each spectrum, the CANDELS ID of the source and the targeted line are given in the top left corner, the black line is the continuum-subtracted and primary beam-corrected spectrum, the gray line is the rms noise per channel, the vertical dashed green line is the OIR spectroscopic redshift (see Table \ref{tab:sample}), and the red curve is the best-fitting Gaussian. Blue and orange shading show the line channels ($\pm2\sigma$ from the center) for CO and [C\,{\sc i}] emission, respectively. All spectra show a bandwidth of 2500\,km\,s$^{-1}$ centered on the Gaussian. To the right of each spectrum is a $12''\times12''$ thumbnail showing the cleaned line-only images produced from these channels, with the NOEMA synthesized beam shape shown in the bottom right corner. Each thumbnail is centered on the CANDELS position, which is shown by the red circle and cross. The 2D S/N of these images is given in the top left corner of each thumbnail. Contours are at 2, 4, 6, 8, 10, and 12$\sigma$.}
	\label{fig:spec}
\end{figure*}

We obtained secure detections of 7/8 of our targeted CO lines (see Columns~(2) and (3) of Table~\ref{tab:lines}), as well as a secure detection of [C\,{\sc i}](1--0) in GN-3987 and a marginal [C\,{\sc i}] detection in GN-8051. In Figure~\ref{fig:spec}, we show the baseline-subtracted, primary beam-corrected spectra and their best-fit Gaussian profiles. Each panel shows a total bandwidth of 2500\,km\,s$^{-1}$ centered on the best-fit redshift. We highlight the line channels (those within $\pm2\sigma$ of the best-fit redshift) in blue for CO and in orange for [C\,{\sc i}]. Our CO detections have full widths at half-maximum (${\rm FWHM} \equiv 2.355\,\sigma$) in the range 263--963 km\,s$^{-1}$, slightly above the range of CO lines in the ASPECS-LP sample (FWHM = 40--609 km\,s$^{-1}$; \citetalias{Aravena19}). In Columns~(4) and (5) of Table~\ref{tab:lines}, we list the best-fit redshifts and FWHM of the detected lines. The uncertainties on the FWHM are the 16th to 84th percentile values of the distributions of these parameters from our MC simulations. 

The CO and [C\,{\sc i}] redshifts from our NOEMA emission line detections are consistent with their existing OIR spectroscopic redshifts, which are shown by the vertical dashed lines in Figure \ref{fig:spec}. Alongside each spectrum, we also show the line-only image for the detected line (see Section \ref{sec:line2d}), and mark the CANDELS position with the red circle and cross. All of our line detections are coincident with the red-selected CANDELS source position within $<1''$.

Notably, in GN-3987, we detect both CO(4--3) and [C\,{\sc i}](1--0) at $z = 2.215$, rather than the $z = 2.203$ of the ``candidate'' CO(3--2) line associated with its SCUBA-2 source \citep{Bothwell13}. Both redshifts are consistent with the highly uncertain \citet{Kodra23} redshift of $z = 2.21\pm0.03$, which was fit to broad polycyclic aromatic hydrocarbon features in the MIR spectrum \citep{Pope08}. However whereas both of our line detections have positions coincident with the red CANDELS galaxy (and a \citealt{Owen18} radio source), the candidate CO(3--2) detection was located $3.6''$ away. We extracted 1D spectra from our NOEMA data at the \citet{Bothwell13} position to search for CO(4--3) and [C\,{\sc i}](1--0) at $z = 2.203$, but find no evidence of either line, despite our deeper observations. We therefore adopt our new redshift of $z = 2.215$ for GN-3987 for the remainder of this work.

\subsubsection{Integrated Fluxes and Luminosities}
\label{sec:line_flux}

We measure integrated CO fluxes of $I_{\rm CO} = 0.36-3.27\,{\rm Jy\,km\,s^{-1}}$ in the detected emission lines, with five of our seven detections having $I_{\rm CO} < 1\,{\rm Jy\,km\,s^{-1}}$. We convert the integrated fluxes to line luminosities, in brightness temperature units, via Equation 3 from \citet{Solomon05}:
\begin{equation}
	L'_{\rm line}~[{\rm K\,km\,s^{-1}\,pc^2}] = \frac{3.25 \times 10^7\, I_{\rm line} \, D_{\rm L}^2}{ \bar \nu^2 \, (1 + z)^3} \,,\label{eqn:LCO}
\end{equation}
where $I_{\rm line}$ is the integrated line flux in Jy km s$^{-1}$, $\bar \nu$ is the line sky frequency in GHz, and $D_{\rm L}$ is the luminosity distance in Mpc, calculated using the redshift of the CO or [C\,{\sc i}] line. We obtain line luminosities of $L'_{\rm line} = (0.5-4.9) \times 10^{10}\,{\rm K\,km\,s^{-1}\,pc^2}$, or $L'_{\rm line} = (0.6-4.9) \times 10^{10}\,{\rm K\,km\,s^{-1}\,pc^2}$ for just the CO detections. 
In Columns (6) and (7) of Table~\ref{tab:lines}, we give the integrated fluxes and line luminosities for each emission line.

To measure upper limits to the integrated fluxes of the two undetected lines---CO(4--3) in GN-19876 and [C\,{\sc i}] in GN-23490---we generated continuum-subtracted, line-only datasets for these sources assuming a fixed redshift and FWHM, rather than using the low-S/N Gaussian fits. For the [C\,{\sc i}](1--0) line in GN-23490, we used the redshift and FWHM of the detected CO(4--3) line ($z = 2.319$ and ${\rm FWHM} = 263\,{\rm km\,s^{-1}}$). For GN-19876, since there is no other line to reference for FWHM, we assumed the OIR redshift of $z = 2.466$ and ${\rm FWHM} = 409\,{\rm km\,s^{-1}}$, which is the median FWHM of the detected CO(4--3) lines in our sample. We then imaged these and measured the rms in the image plane in the same way that we did for the detected emission lines. We took the upper limits to the integrated fluxes to be
\begin{equation}
    I_{\rm line} \leq 4 \cdot {\rm rms} \cdot 4\sigma \,.
\end{equation}
Here, $\sigma$ is the assumed Gaussian line width. 

In Table \ref{tab:lines}, we provide the upper limits to the integrated fluxes and corresponding line luminosities for the two undetected lines. We note that in addition to the errors listed in this Table, all of our NOEMA measurements are subject to systematic uncertainties in absolute flux calibration of $\sim$10\%.

\subsection{2 mm Continuum}
\label{sec:continuum}

Following the identification of emission lines in our NOEMA images, we generated continuum-only $uv$ data and images for each sideband by averaging together all of the line-free channels (i.e., those falling outside of $\pm2\sigma$ from the best-fit line center). We then measured the fluxes and positions of continuum emission in the $uv$ plane, as well as imaging and deconvolving the continuum data, using the same procedure we used to measure and image our line-only images (Section \ref{sec:line2d}). For sidebands with off-axis sources (see Section \ref{sec:other_sources}), we fit point sources to both positions simultaneously, giving \texttt{uv\_fit} initial guesses for the source position within $\pm2''$ of each CANDELS position.

We classified a continuum source as a detection if the best-fit point source flux exceeded both 4 times the rms in that image and the absolute value of the most negative pixel value within the extent of the NOEMA primary beam. The S/N of these negative ``sources'' ranged from $-2.6$ to $-4.4$, with stronger negative peaks in the C and C/D configuration images, as expected given the higher number of independent synthesized beams that cover the primary beam area at higher spatial resolution.

Using these criteria, we detected continuum emission in both sidebands in GN-3835, GN-3987, and GN-8051, as well as in the USB of GN-19084. We report these values, along with the central frequency of the sideband in which they were measured, in Table \ref{tab:continuum}. For all of the sidebands without continuum detections, we provide upper limits (either $4 \times {\rm rms_{2D}}$ or the inverse of the most negative peak, whichever is greater) in Table \ref{tab:continuum}.

\begin{table}[ht]
	\centering
	\caption{NOEMA Continuum Fluxes and Upper Limits}
	\begin{tabularx}{\linewidth}{rrrrrrr}
		\hline \hline
              & \multicolumn{2}{c}{LSB} & \multicolumn{2}{c}{USB} \\
		ID    & $\nu$    & Flux & $\nu$ & Flux \\
		       & (GHz)    & ($\mu$Jy\,beam$^{-1}$) & (GHz) & ($\mu$Jy\,beam$^{-1}$) \\
		\hline
        \multicolumn{5}{l}{\textit{Detections}}             \\
         3835 & 139.3 & $197 \pm 41$ & 154.7 & $255 \pm 70$ \\
         3987 & 141.3 & $ 94 \pm 18$ & 156.7 & $177 \pm 22$ \\
         8051 & 136.3 & $173 \pm 25$ & 151.7 & $261 \pm 33$ \\
        19084 & ---   & ---          & 151.7 & $539 \pm 99$ \\
        \multicolumn{5}{l}{\textit{Non-detections}}         \\
        19084 & 136.3 & $<336      $ & ---   & ---          \\
         7630 & 136.3 & $< 69      $ & 151.7 & $< 72      $ \\
        10283 & 139.3 & $<281      $ & 154.7 & $<286      $ \\
        10601 & 131.3 & $< 74      $ & 146.7 & $< 84      $ \\
        13722 & 141.3 & $< 52      $ & 156.7 & $< 62      $ \\
        19876 & 136.3 & $<175      $ & 151.7 & $<166      $ \\
        23490 & 136.3 & $<148      $ & 151.7 & $<149      $ \\
		\hline
	\end{tabularx}
	\label{tab:continuum}
\end{table}

\section{Physical Properties}
\label{sec:physical}

\subsection{Molecular Gas Masses from CO Line Luminosities}
\label{sec:gas}

We derived molecular gas mass estimates for the eight galaxies in our NOEMA sample---the seven CO detections and the CO(4--3) upper limit for GN-19876---from their CO line luminosities. We first converted the luminosities of the observed CO transitions to those of the fundamental CO(1--0) transitions by assuming the median values of $r_{31} \equiv L'_{\rm CO(3-2)} / L'_{\rm CO(1-0)} = 0.63\pm0.12$ and $r_{41} \equiv L'_{\rm CO(4-3)}/L'_{\rm CO(1-0)} = 0.34\pm0.04$ from \citetalias{Birkin21}. We give these CO(1--0) luminosities in Column~(8) of Table~\ref{tab:lines}. We note that the line ratios for individual DSFGs can vary significantly---for example, \citet{Riechers20} measured $r_{\rm 31} = 0.23-1.56$ in a sample of six ASPECS galaxies with follow-up CO(1--0) detections using the VLA; and \citetalias{Aravena19} used ratios of $r_{31} = 0.42\pm0.07$ and $r_{41} = 0.31\pm0.06$, which were derived from a survey of $BzK$-selected massive main sequence galaxies at $z = 1.5$ \citep{Daddi15}.

The galaxies' CO(1--0) luminosities are then related to the molecular gas by 
\begin{equation}
	M_{\rm mol} = 1.36 \, \alpha_{\rm CO} \, L'_{\rm CO(1-0)}, \label{eqn:mmol}
\end{equation}
where the factor of 1.36 accounts for a 10\% Helium fraction, and the conversion factor $\alpha_{\rm CO}$ has units of $M_\odot \, ({\rm K\,km\,s^{-1}\,pc^2})^{-1}$. The CO conversion factor varies based on several factors, especially metallicity \citep[see, e.g., the review by][]{Bolatto13}. In this work, we assume a fiducial value of $\alpha_{\rm CO} = 3.6$ \citep{Daddi10}, which is consistent with the analysis of ASPECS-LP \citepalias{Aravena19}, though we note that other studies have adopted lower values \citep[e.g., $\alpha_{\rm CO} = 1.0$,][]{Bothwell13}. 

Our measured gas masses using Equation \ref{eqn:mmol} are in the range $\log M_{\rm mol} / M_\odot = 10.8-11.9$ dex, with a median of 11.2 dex. From the CO(4--3) upper limit in GN-19876, we inferred an upper limit of $\log M_{\rm  mol} / M_\odot < 11.2$ dex. In Column~(9) of Table~\ref{tab:lines}, we list the gas mass or upper limit that we derived from each targeted CO line.

\subsubsection{Neutral Carbon-based Gas Masses}
\label{sec:ci_gas}

Molecular gas can also be inferred from the [C\,{\sc i}](1--0) line, and we measure [C\,{\sc i}]-based gas masses for the three galaxies with spectral coverage of this line. We assume a line luminosity (or conversion factor) ratio of $L'_{\rm CO(1-0)} / L'_{\rm [CI](1-0)} = \alpha_{\rm [CI]}/\alpha_{\rm CO} = 4.5\pm0.2$, which is the median value for local ULIRGs \citep{Jiao17} and similar to $\alpha_{\rm [CI]}/\alpha_{\rm CO} \sim 4.5$ found for $z > 2$ DSFGs (e.g., \citetalias{Birkin21}; \citealp{Dunne22}). In Columns~(8) and (9) of Table~\ref{tab:lines}, we give these CO(1-0) luminosities or upper limits, and the gas masses derived from them. 

The two galaxies with both CO(4--3) and [C\,{\sc i}](1--0) in our spectral coverage are GN-23490 and GN-3987. GN-3987 has strong detections of both lines. We inferred gas masses of $1.5\times10^{11}$ and $1.1\times10^{11}\,M_\odot$ from CO(4--3) and [C\,{\sc i}](1--0), respectively, which are nearly in agreement. We do not detect the [C\,{\sc i}](1--0) line in GN-23490, and the gas mass inferred from the CO(4--3) detection is 2.3 times higher than the 4$\sigma$ upper limit on the [C\,{\sc i}]-derived mass. High CO(4--3) to [C\,{\sc i}](1--0) ratios can indicate higher gas densities, which allow the CO to be shielded against photodissociation by the interstellar radiation field \citep{Kaufman99}. We note that GN-23490 has the lowest FWHM of any CO detection in our sample, which may support a dense, central gas reservoir scenario.

For the purposes of calculating derived properties that depend on $M_{\rm mol}$ (Section \ref{sec:discussion}), we exclusively use the CO-based $M_{\rm mol}$ measurements where available, since there are many more CO-based gas masses in the literature than [C\,{\sc i}]-based masses. This allows us to minimize systematic uncertainties when comparing the gas properties of our sample to those of other CO surveys and discussing strategies for future surveys of the faint DSFG population.

\subsection{SED Fitting}

To estimate the physical properties of the DSFGs that host the molecular gas reservoirs that we measured in Section~\ref{sec:gas}, we fit the DSFGs' SEDs using two different fitting codes. To measure the dust attenuation and stellar masses, we used BAGPIPES \citep[Bayesian Analysis of Galaxies for Physical Inference and Parameter EStimation;][]{Carnall18}, a stellar population synthesis (SPS)-based code, on the full UV-to-mm photometry of our sample. Second, we performed FIR-only modified blackbody (MBB) fits \citep{McKay23} to obtain SFH-independent measurements of our galaxies' SFRs. We describe the input photometry, our fitting procedure, and the resulting estimated physical properties in the remainder of this Section.

\subsubsection{Input Photometry}

We performed our SED fits using the CANDELS UV-to-NIR photometry from \citet{Barro19}, supplemented with longer wavelength data from the literature. The CANDELS photometry includes ground-based $U$, $K$, and $K_s$ imaging, five HST/ACS and four WFC3-IR bands, and all four Spitzer/IRAC bands. At longer wavelengths, we used Spitzer/MIPS 24\,$\mu$m, Herschel/PACS 100 and 160\,$\mu$m, and Herschel/SPIRE 250 and 350\,$\mu$m fluxes from the GOODS-Herschel survey \citep{Elbaz11}, with all nine faint DSFGs in this work having a match with an angular separation of $<1''$; 450 and 850\,$\mu$m SCUBA-2 fluxes from SUPER GOODS; 1.3\,mm fluxes or upper limits from the NIKA-2 Cosmological Legacy Survey \citep[][]{Bing23}; and our NOEMA 2\,mm fluxes and upper limits from Table~\ref{tab:continuum}. If a source was not found in any of the above catalogs, then we measured an upper limit by taking the noise value at the position of the CANDELS galaxy.

\setlength{\tabcolsep}{7pt}
\begin{table*}
	\centering
	\caption{Physical Properties}
    \footnotesize
	\begin{tabularx}{\linewidth}{cccccccccc}
		\hline \hline
		ID    &  $A_{\rm V}$           & $\log \frac{M_\star}{M_\odot}$ & SFR$_{\rm SED}$ & $\log \frac{L_{\rm IR}}{L_\odot}$ & SFR$_{\rm KE12}$ & $\Delta$MS          & $\log \frac{M_{\rm mol}}{M_\odot}$ & $t_{\rm dep}$ & $\mu_{\rm mol}$ \\
              & (mag)                  &                         & ($M_\odot\,{\rm yr}^{-1}$) &                 & ($M_\odot\,{\rm yr}^{-1}$) & (dex)                   &                                    & (Gyr)         &               \\
         (1)  & (2)                    & (3)                     & (4)                & (5)                     & (6)                        & (7)                     & (8)                                & (9)           & (10)          \\
        \hline
        10283 & $1.69^{+0.04}_{-0.04}$ & $11.12^{+0.01}_{-0.02}$ & $155^{+21}_{-19}$  & $12.56^{+0.02}_{-0.02}$ & $517^{+21}_{-20}$          &  $0.78^{+0.02}_{-0.04}$ & $11.35^{+0.08}_{-0.10}$            & $0.44\pm0.09$ & $1.70\pm0.35$ \\
        13722 & $1.89^{+0.04}_{-0.04}$ & $11.06^{+0.03}_{-0.03}$ & $87^{+5}_{-6}$     & $12.00^{+0.04}_{-0.04}$ & $142^{+12}_{-12}$          &  $0.15^{+0.04}_{-0.06}$ & $10.79^{+0.08}_{-0.10}$            & $0.43\pm0.09$ & $0.53\pm0.11$ \\
        10601 & $1.38^{+0.09}_{-0.09}$ & $11.52^{+0.02}_{-0.02}$ & $98^{+64}_{-31}$   & $12.43^{+0.03}_{-0.03}$ & $382^{+23}_{-22}$          &  $0.17^{+0.03}_{-0.05}$ & $10.91^{+0.09}_{-0.11}$            & $0.21\pm0.05$ & $0.24\pm0.05$ \\
        3987  & $2.85^{+0.12}_{-0.10}$ & $11.47^{+0.05}_{-0.04}$ & $136^{+15}_{-14}$  & $12.26^{+0.05}_{-0.05}$ & $256^{+28}_{-28}$          & $-0.16^{+0.05}_{-0.08}$ & $11.19^{+0.06}_{-0.07}$            & $0.60\pm0.09$ & $0.50\pm0.07$ \\
        3835  & $1.59^{+0.12}_{-0.05}$ & $10.95^{+0.03}_{-0.04}$ & $67^{+13}_{-7}$    & $12.68^{+0.02}_{-0.02}$ & $668^{+35}_{-35}$          &  $0.63^{+0.02}_{-0.05}$ & $11.85^{+0.08}_{-0.09}$            & $1.07\pm0.21$ & $7.95\pm1.54$ \\
        23490 & $1.77^{+0.07}_{-0.08}$ & $11.44^{+0.03}_{-0.03}$ & $138^{+15}_{-14}$  & $12.23^{+0.05}_{-0.06}$ & $239^{+31}_{-32}$          & $-0.19^{+0.05}_{-0.08}$ & $11.23^{+0.07}_{-0.08}$            & $0.72\pm0.12$ & $0.62\pm0.10$ \\
        7630  & $1.65^{+0.10}_{-0.11}$ & $10.40^{+0.06}_{-0.08}$ & $56^{+15}_{-13}$   & $12.16^{+0.08}_{-0.09}$ & $206^{+43}_{-41}$          &  $0.52^{+0.09}_{-0.11}$ & $10.93^{+0.08}_{-0.10}$            & $0.41\pm0.08$ & $3.36\pm0.67$ \\
        \hline
        19876 & $1.47^{+0.05}_{-0.04}$ & $11.31^{+0.02}_{-0.02}$ & $135^{+17}_{-13}$  & $12.69^{+0.03}_{-0.03}$ & $687^{+45}_{-41}$          &  $0.33^{+0.03}_{-0.04}$ & $<$11.20                           & $<$0.25       & $<$0.77       \\
        8051* & $5.00^{+0.34}_{-0.64}$ & $11.34^{+0.06}_{-0.04}$ & $177^{+69}_{-177}$ & $12.42^{+0.05}_{-0.07}$ & $368^{+51}_{-51}$          & $-0.03^{+0.06}_{-0.11}$ & $11.12^{+0.10}_{-0.11}$            & $0.26\pm0.06$ & $0.56\pm0.13$ \\
		\hline
	\end{tabularx}
	\begin{tablenotes}
    \item*The 68\% confidence interval on BAGPIPES fits of GN-8051's SED using a \citet{CharlotFall00} law includes quiescent galaxy solutions, which are inconsistent with the $L_{\rm IR}$ and $M_{\rm mol}$ measurements. Fitting the same galaxy with a \citet{Calzetti00} law returns $A_{\rm V} = 2.87$ and $\log M_\star/M_\odot = 10.12$. Assuming this stellar mass implies that GN-8051 is a strongly starbursting galaxy, with $\Delta{\rm MS} = 0.96$ dex and $\mu_{\rm mol} \simeq 10$.
    \end{tablenotes}
    \label{tab:properties}
\end{table*}

\subsubsection{BAGPIPES: $A_V$, Stellar Mass, $L_{IR}$, and SFR}
\label{sec:bagpipes}

\begin{figure*}
	\centering
	\includegraphics[width=\textwidth]{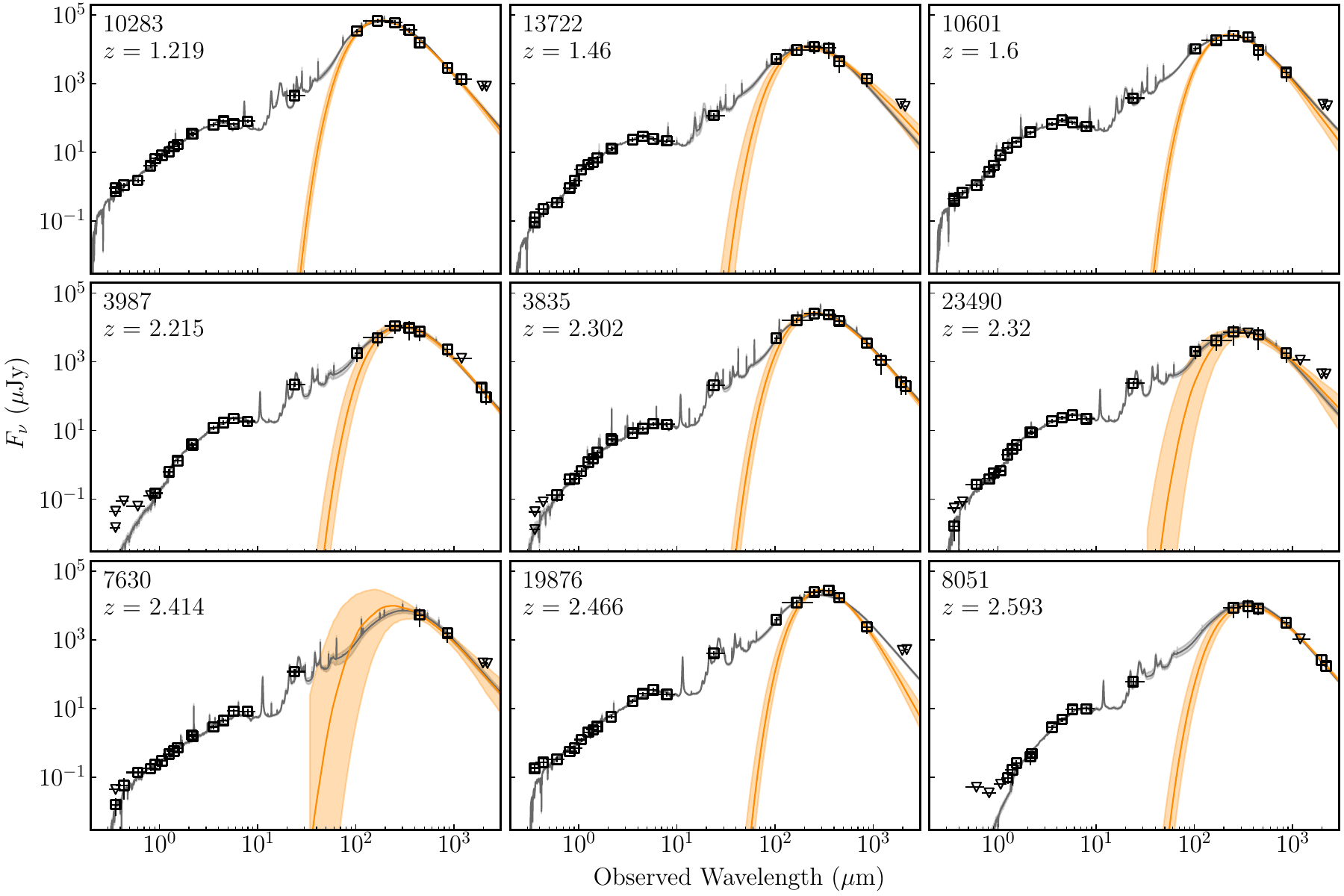}
	\caption{SEDs of the the nine faint DSFGs with spectroscopic redshifts. The CANDELS ID and redshift are shown in the top-left corner of each panel. The gray curve and shaded region in each panel show the median posterior SED and the 16th to 84th percentile range, respectively. The orange curve and shaded region show the median optically-thin MBB SED and the 16th to 84th percentile range, respectively. The square points indicate detections, while the triangles indicate $3\sigma$ upper limits. All vertical errorbars show $3\times$ the measured error.}
    \label{fig:sed}
\end{figure*}

We ran our BAGPIPES UV-to-mm fits using the stellar population models of \citet{Bruzual03} and a physically motivated silicate--graphite--PAH dust emission model from \citet{Draine07}. We assumed a delayed SFH with an optional starburst, with logarithmic priors on the age and decay time, $\tau$, of both SFH components, an ionization parameter of $\log U$ between $-3.5$ and $-2$, and a wide range of PAH emission fractions and metallicities. We assumed a \citet{CharlotFall00} dust attenuation curve. These models and parameters were selected to resemble those of the Multi-wavelength Analysis of Galaxy PHYSical properties \citep[MAGPHYS;][]{daCunha08} code, which was used to measure the properties of the CO-detected DSFGs in \citetalias{Aravena19} and \citetalias{Birkin21}. In Figure~\ref{fig:sed}, we show the resulting SEDs for each galaxy, and in Table~\ref{tab:properties}, we give the medians and 68\% CL errors on $A_{\rm V}$, $\log M_\star/M_\odot$, and $\log L_{\rm IR}/L_\odot$ from our fits.

In Table \ref{tab:properties}, we also provide two different measures of the SFR. SFR$_{\rm SED}$ is the 100\,Myr average of the fitted SFH in BAGPIPES, and SFR$_{\rm KE12}$ is calculated from $L_{\rm IR}$ using the relation:
\begin{equation}
    {\rm SFR_{KE12}}~[M_\odot\,{\rm yr}^{-1}] = 1.41\times10^{-10}~L_{\rm IR}~[L_\odot] \label{eqn:ir_sfr} \,,
\end{equation}
from \citet[][hereafter KE12]{Kennicutt12}, which is scaled to a \citet{Chabrier03} initial mass function (IMF). The latter measurement is also sensitive to star formation over $\sim$100 Myr timescales. 

Our sample has ${\rm SFR_{KE12}} = 140-690\,M_\odot\,{\rm yr^{-1}}$, with a median of 370 $M_\odot\,{\rm yr^{-1}}$. For all nine of our galaxies, these are higher than the SFRs returned by BAGPIPES, which span a range of ${\rm SFR_{SED}} = 60-180\,M_\odot\,{\rm yr}^{-1}$, with a median ${\rm SFR_{SED}}$ of 135 $M_\odot\,{\rm yr}^{-1}$. The median ratio of these two SFR measures in our sample is $\langle {\rm SFR_{KE12} / SFR_{SED}}\rangle = 2.1$ (0.32 dex), and the largest individual ratio is a factor of 10, in GN-3835. This median offset is large, but it is consistent with uncertainties in SFRs generated by SPS-based SED fitting codes ($\sim$0.1--0.5 dex; \citealp{Pacifici15}). The individual offsets are also consistent with findings that SED fits that use parametric SFHs (such as MAGPHYS and our BAGPIPES fits) can underpredict SFRs for individual galaxies by up to an order of magnitude \citep{Carnall19} compared to the \citetalias{Kennicutt12} relations. 

For the remainder of this work, we use ${\rm SFR_{KE12}}$ unless otherwise stated, since this minimizes differences between measured SFRs due to SED fitting assumptions, and it allows us to compare easily the properties of our faint DSFG sample to those of other samples, as most authors provide $L_{\rm IR}$.

We caution that for the serendipitously-detected GN-8051, which is the OIR-faintest (F160W = 25.4 mag) and reddest galaxy in this work (see Figure \ref{fig:selection}), the physical properties that we infer from our BAGPIPES fits are likely biased due to our use of a \citet{CharlotFall00} attenuation law. The 68\% confidence interval on SFR$_{\rm SED}$ that is returned by these fits extends to quiescent solutions with $<1M_\odot\,{\rm yr}^{-1}$, which is inconsistent with the massive gas reservoir of $\log M_{\rm mol} / M_\odot > 11$ that we inferred from our [C\,{\sc i}](1--0) detection, and the high $L_{\rm IR}$, which implies ${\rm SFR_{KE12}} = 368\pm51\,M_\odot\,{\rm yr}^{-1}$. When we refit GN-8051's SED with a \citet{Calzetti00} modified starburst extinction law---keeping all other input models the same---we obtained $A_{\rm V} = 2.87^{+0.15}_{-0.13}$, with a similar ${\rm SFR_{SED}} = 162^{+30}_{-28}\,M_\odot\,{\rm yr}^{-1}$ and a much lower stellar mass of only $\log M_\star/M_\odot = 10.12$.

\setlength{\tabcolsep}{9pt}
\begin{table}
	\centering
	\caption{Modified Blackbody SED Fits}
	\begin{tabularx}{\linewidth}{rcccc}
		\hline \hline
		ID & $T$ & $\log \frac{L_{\rm IR}}{L_\odot}$ & SFR$_{\rm KE12}$ & $\beta$\\
		 & (K) &  & (M$_\sun$\,yr$^{-1}$) &  \\
		(1) & (2) & (3) & (4) & (5) \\
		\hline
10283 & $41^{+5}_{-4}$ & $12.62^{+0.06}_{-0.05}$ & $582^{+84}_{-69}$ & $1.5^{+0.2}_{-0.2}$ \\
13722 & $50^{+13}_{-9}$ & $12.01^{+0.08}_{-0.07}$ & $144^{+30}_{-20}$ & $0.7^{+0.5}_{-0.4}$ \\
10601 & $43^{+8}_{-7}$ & $12.34^{+0.05}_{-0.04}$ & $312^{+38}_{-31}$ & $1.2^{+0.5}_{-0.4}$ \\
3987 & $39^{+38}_{-16}$ & $12.39^{+0.77}_{-0.47}$ & $349^{+1698}_{-231}$ & $1.6^{+0.6}_{-0.4}$ \\
3835 & $37^{+5}_{-4}$ & $12.63^{+0.08}_{-0.07}$ & $596^{+116}_{-85}$ & $1.7^{+0.3}_{-0.2}$ \\
23490 & $55^{+41}_{-32}$ & $12.51^{+0.68}_{-0.67}$ & $456^{+1748}_{-359}$ & $1.2^{+1.3}_{-0.7}$ \\
7630 & $50^{+44}_{-27}$ & $12.42^{+0.73}_{-0.61}$ & $367^{+1600}_{-278}$ & $1.4^{+1.4}_{-0.7}$ \\
19876 & $33^{+9}_{-6}$ & $12.63^{+0.09}_{-0.07}$ & $602^{+140}_{-86}$ & $2.4^{+0.7}_{-0.7}$ \\
8051 & $33^{+22}_{-9}$ & $12.30^{+0.47}_{-0.24}$ & $279^{+549}_{-119}$ & $1.5^{+0.4}_{-0.5}$ \\
		\hline
	\end{tabularx}
    \label{tab:mbbfit}
\end{table}
\setlength{\tabcolsep}{6pt}

\subsubsection{FIR-Only Modified Blackbody Fits}
\label{sec:dust_sed}

In addition to the BAGPIPES fits, we performed FIR-only SED fits on our faint DSFG sample following the procedure of \citet{McKay23}. These fits were motivated by the fact that the shortest wavelength dust emission ($\lambda_{\rm rest} \leq 40\,\mu$m) can be impacted by AGN heating, whereas the longer-wavelength emission arises almost entirely from star formation-heated dust in the interstellar medium (ISM) \citepalias[e.g.,][]{Kennicutt12}.

We fit the observed-frame 160\,$\mu$m to 2\,mm fluxes of each galaxy with a MBB template using the Python-based Markov Chain Monte Carlo (MCMC) code \texttt{emcee} \citep{Foreman-Mackey13}. The model has three free parameters: the blackbody temperature, $T$, the dust emissivity, $\beta$, and the overall normalization. The MBB model and fitting procedure\footnote{\url{https://mbb.readthedocs.io/en/latest/}} are described in further detail in \citet{McKay23}. In Figure~\ref{fig:sed}, we show the best-fit MBB SED and 16th to 84th percentile flux interval for each of the nine faint DSFGs in orange, and in Table~\ref{tab:mbbfit}, we provide the median and 16th to 84th percentile posterior values of $T$ and $\beta$. We measure blackbody temperatures\footnote{This is different from the characteristic dust temperature $T_{\rm d}$, which is approximated via Wien's law, $T_{\rm d} = (b/\lambda_{\rm peak})^{1/0.9}$, from which we find our DSFGs have $T_{\rm d} = 29-59$ K.} in the range $T = 33-55$ K and emissivities in the range $\beta = 0.7-2.4$. Both of these are consistent with the typical observed properties of DSFGs in this redshift and $S_{\rm 850 \mu m}$ range \citep{McKay23}, though the errors on $T$ are large for galaxies with fewer FIR detections.

As can be seen in Figure~\ref{fig:sed}, the optically thin MBB-only SEDs decline very rapidly in the MIR compared to the BAGPIPES SEDs. To correct for the missing MIR contribution to the 8-1000\,$\mu$m total IR luminosity, we multiply the integrated luminosities from our MBB fits by a factor of 1.3, following \citetalias{Kennicutt12}. We provide the corrected $L_{\rm IR}$ for each galaxy in Table~\ref{tab:mbbfit}. We find a range of $\log L_{\rm IR} / {\rm L}_\odot = 12.0-12.6$, with a median of 12.4. These luminosities correspond to ${\rm SFR_{KE12}} \simeq 140-600\,{\rm M}_\odot\,{\rm yr}^{-1}$, with a median of 370\,${\rm M}_\odot\,{\rm yr}^{-1}$. 

Comparing the infrared luminosities inferred from these MBB fits to those from BAGPIPES, we find a median difference of $\log L_{\rm IR}^{\rm bagpipes}/L_{\rm IR}^{\rm mbb} = -0.01$\,dex. The median log difference in $L_{\rm IR}$ for just the four DSFGs containing X-ray AGN (GN-3835, 10283, 10601, and 19876) was slightly higher (+0.05\,dex) than for the non-AGN (--0.14\,dex), but all of these are smaller than the median measured error on $L_{\rm IR}$ from either fit, suggesting that AGN do not strongly impact the $L_{\rm IR}$ or SFR measurements for our sample. 

\section{Discussion}
\label{sec:discussion}

\begin{figure*}[tb]
   \centering
   \includegraphics[width=0.8\linewidth]{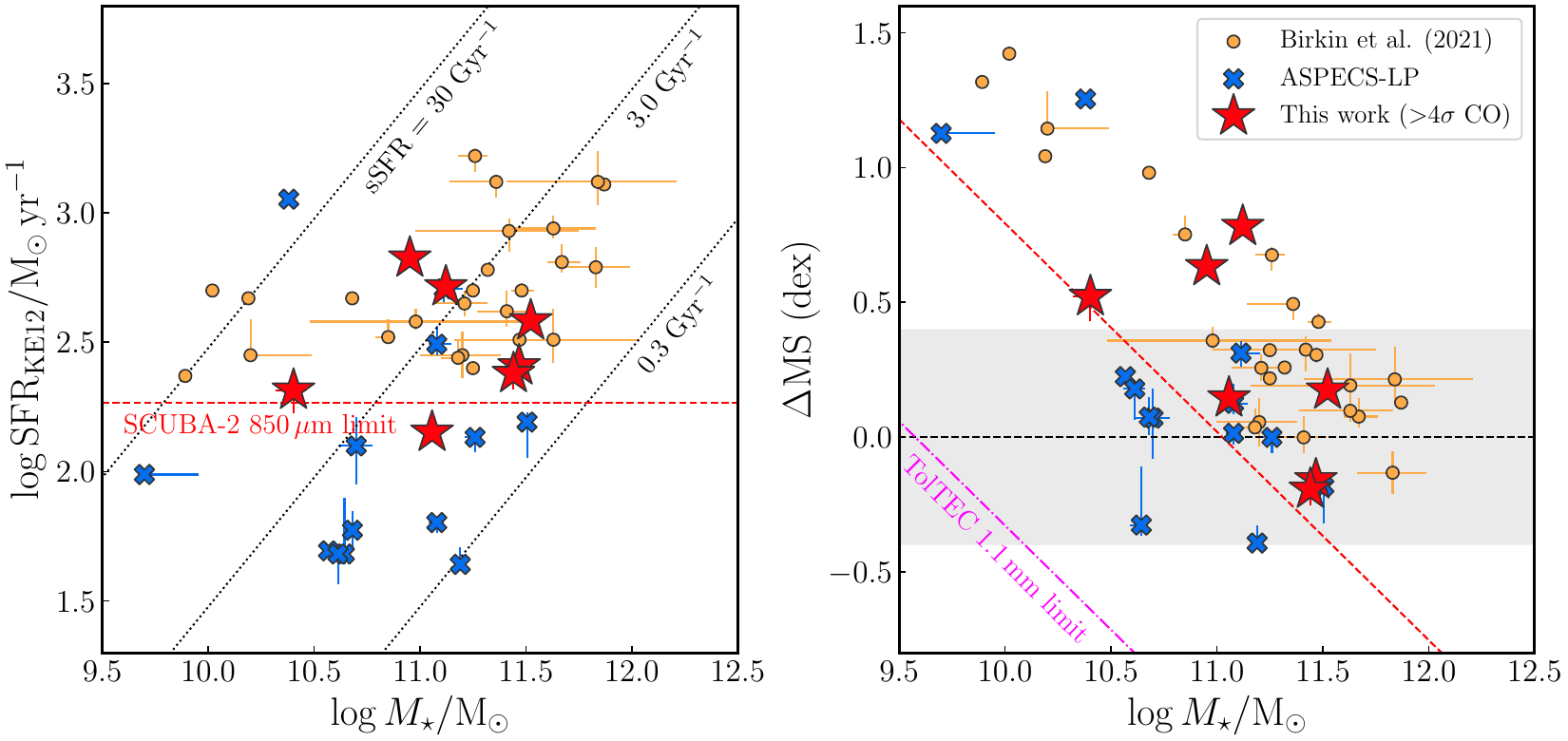}
   \caption{
   CO-detected DSFGs at $z = 1-3$ from this work (red stars), \citet{Birkin21}
   (orange circles), and ASPECS-LP (blue X's). All SFRs were calculated using Equation~\ref{eqn:ir_sfr}. \textit{Left:} SFR vs. stellar mass. Lines of fixed ${\rm sSFR = SFR} / M_\star$ are shown. The approximate SFR corresponding to the confusion limit of SCUBA-2 at 850\,$\mu$m ($185\,{\rm M_\odot\,yr^{-1}}$) is shown by the red dashed line. \textit{Right:} Main sequence offset ($\Delta$MS) vs. stellar mass. The black dotted line indicates the \citet{Speagle14} SFMS, and a region of $\pm0.4$\,dex around this is shaded grey. The same SFR limit as in the left panel is plotted for $z = 2.5$; low-mass ($M_\star \lesssim 10^{11}\,{\rm M}_\odot$), main-sequence galaxies are excluded from the follow-up CO samples due to the SFR limits of the parent submm samples. The limiting luminosity for TolTEC at 1.1\,mm of $L_{\rm IR} = 10^{11}\,{\rm L}_\odot$ ($14\,{\rm M_\odot\,yr^{-1}}$; \citealp{NavaMoreno24}), which will enable the selection of galaxies on the SFMS at lower masses, is shown by the magenta dot-dashed line. }
   \label{fig:main_sequence}
\end{figure*}

We obtained new CO detections for seven faint DSFGs in the GOODS-N, six of which are the first CO detections ever made for these galaxies. To contextualize the results of our survey, we compare the physical properties of our new CO sample (Table~\ref{tab:properties}) with those of CO-detected galaxies at $z = 1-3$ from two other large interferometric surveys, namely:

\begin{enumerate}
    \item The \citetalias{Birkin21} CO detections from ALMA and NOEMA follow-up of ALMA continuum sources, down to $S_{\rm 870\mu m} = 2$\,mJy, obtained using a combination of wide-bandwidth linescans (for galaxies with unknown redshifts) and direct targeting of CO at known frequencies, as in this work.
    \item CO detections of mostly faint DSFGs from the ASPECS-LP survey \citepalias{Aravena19}, which obtained extremely deep linescans across 5\,arcmin$^2$ of the Hubble Ultra-Deep Field (HUDF) with contiguous frequency coverage across ALMA Band~3.
\end{enumerate}
In addition to being aimed at detecting CO in relatively faint galaxies, both of these works measured their galaxies' physical properties using the MAGPHYS code, which employs similar SED-fitting assumptions as this work, as we discussed in Section \ref{sec:physical}. To make accurate comparisons across the three samples, we calculated IR-based SFRs for the \citetalias{Birkin21} and \citetalias{Aravena19} galaxies using their $L_{\rm IR}$ values and Equation \ref{eqn:ir_sfr}. We limit our comparison to galaxies with $>$4$\sigma$ CO detections at $z = 1-3$, i.e., our seven CO-detected faint DSFGs, 25/50 galaxies from \citetalias{Birkin21}, and 13/16 CO-selected galaxies from ASPECS-LP. 

We show the SFRs and stellar masses of all three samples in Figure \ref{fig:main_sequence}, with our CO detections shown by the red stars, \citetalias{Birkin21} galaxies by the orange circles, and ASPECS-LP galaxies by the blue X's. Our CO detections have slightly lower SFRs than the \citetalias{Birkin21} galaxies, as expected given our sample's lower submm flux limits ($S_{\rm 850 \mu m} \geq 1.4$\,mJy vs. $S_{\rm 870 \mu m} \geq 2.0$\,mJy). The two faintest galaxies in our sample, GN-13722 and GN-7630, have lower SFRs than the entire \citetalias{Birkin21} sample (142 and 206\,${\rm M}_\odot\,{\rm yr}^{-1}$, respecitvely). The ASPECS-LP galaxies extend to much lower SFRs, down to $\sim$40\,${\rm M_\odot\,yr^{-1}}$. Together, the three samples include 45 CO-detected galaxies spanning roughly two orders of magnitude in $M_\star$ and SFR, allowing us to investigate the relationships between molecular gas and other physical properties for a wide variety of galaxies using similar assumptions.

\subsection{Missing Low-$M_\star$, Main-Sequence DSFGs}
\label{sec:main_sequence}

The SFRs and stellar masses shown in the left panel of Figure \ref{fig:main_sequence} are not tightly correlated, such that the CO detections span a large range of specific star formation rates, ${\rm sSFR} \equiv {\rm SFR} / M_\star$. For reference, we plot black dotted lines of fixed ${\rm sSFR} =$ 0.3, 3.0, and 30\,Gyr$^{-1}$ in this panel. This is consistent with recent work showing that faint ($S_{\rm 850 \mu m} < 2$\,mJy) and bright ($>$2\,mJy) red-selected DSFGs have similar $M_\star$ distributions, despite large differences in SFR. \citep{McKay25}.


This wide sSFR distribution primarily a consequence of the DSFG population comprising both main-sequence galaxies and starbursting galaxies. This can be seen in the right panel of Figure \ref{fig:main_sequence}, where we plot the main sequence offsets and stellar masses of the three samples. The main sequence offset is defined as
\begin{equation}
    \Delta{\rm MS}   \equiv \log \left[ \frac{{\rm SFR}}{{\rm SFR}_{\rm MS}(z,M_\star)} \right] \,, \label{eqn:sfms}
\end{equation}
where ${\rm SFR_{MS}}$ is the SFR of the SFMS at the redshift and stellar mass of the observed galaxy. We calculated $\Delta$MS assuming the best-fit prescription from \citet[][their Equation 28]{Speagle14}, shifted by $+0.03$ dex in $M_\star$ to convert to a \citet{Chabrier03} IMF. 
We selected this fiducial SFMS due to their identical assumed cosmology and calibration to the \citetalias{Kennicutt12} $L$--SFR relations. Critically, this SFMS prescription was also assumed by \citetalias{Tacconi20} to measure molecular gas scaling relations (see Section~\ref{sec:scaling_laws}).

Following \citetalias{Aravena19}, we classify galaxies as being on the SFMS if they have $|\Delta{\rm MS}| \leq 0.4$ dex. Our galaxies have $\Delta{\rm MS} = -0.19$ to $+0.78$\,dex, which we list in Column~(7) of Table~\ref{tab:properties}. Three of the nine galaxies (GN-10283, 3835, and 7630) are starburst galaxies, i.e., above the SFMS, whereas the remaining six\footnote{Including GN-19876 and GN-8051, which were not detected in CO and are not shown in Figure \ref{fig:main_sequence}.} fall on the SFMS. None fall below the SFMS. 

Likewise, all of the ASPECS-LP and \citetalias{Birkin21} galaxies are either on or above the \citet{Speagle14} SFMS. We do not reproduce the finding by \citetalias{Aravena19}, who used SFRs from MAGPHYS and a \citet{Schreiber15} SFMS, that two of the ASPECS-LP galaxies are below the SFMS. It is worth noting that classifying individual galaxies as on, above, or below the SFMS depends on the assumed SFR method and SFMS prescription, highlighting the importance of a self-consistent approach when comparing galaxy samples.

The right panel of Figure \ref{fig:main_sequence} also shows that $\Delta$MS declines with increasing $M_\star$. We interpret this as a selection bias arising from the SFR limits of these surveys: the CO samples from this work and \citetalias{Birkin21} are submm flux-limited, with all galaxies having $S_{\rm 850\mu m}$ (or $S_{\rm 870 \mu m}$) brighter than the SCUBA-2 confusion limit of 1.4 mJy. This corresponds\footnote{Assuming a conversion of ${\rm SFR\,[M_\odot\,yr^{-1}]} = 132\times S_{\rm 850 \mu m}$\,[mJy] \citep{McKay25}.} to an SFR limit of $185\,M_\odot\,{\rm yr}^{-1}$, 
which we show with the red dashed lines in Figure \ref{fig:main_sequence}. The diagonal line in the right panel shows the $\Delta$MS--$M_\star$ relationship corresponding to this fixed SFR for a fixed redshift of $z = 2.5$; because galaxies from submm continuum samples cannot fall below this line, low-mass ($M_\star \lesssim 10^{10.5}\,{\rm M}_\odot$) DSFGs selected in this way are exclusively starbursts. 


\subsection{Comparison to Molecular Gas Scaling Relations}
\label{sec:scaling_laws}

While low-mass, main-sequence galaxies are inaccessible to submm-selected surveys such as this work and \citetalias{Birkin21}, the low-mass, starbursting DSFGs are themselves an interesting population that have been under-represented in previous studies of molecular gas at high redshift.

In particular, relatively few CO detections at $z \gtrsim 1$ with this combination of low $M_\star$ and high $\Delta$MS were included in the samples used to measure molecular gas scaling relations (e.g., \citealp{Genzel15, Tacconi18}; \citetalias{Tacconi20}). These studies fit the depletion timescales,  

\begin{equation}
    t_{\rm dep} \equiv \frac{M_{\rm mol}}{{\rm SFR}}
\end{equation}
and the molecular gas fractions,

\begin{equation}
    \mu_{\rm mol} \equiv \frac{M_{\rm mol}}{M_\star}
\end{equation} 
of galaxies as a function of $z$, $M_\star$, and $\Delta$MS, but the bulk of their CO data are for either $z < 1$ \citep[e.g.,][]{Freundlich19} or main-sequence galaxies \citep[e.g.,][]{Tacconi18}. The few low-mass, starbursting galaxies that were included in these studies have depletion times and gas fractions roughly twice as high as the best-fit scaling relations predict \citep[see][their Figure 7]{Tacconi18}. 

In this subsection, we compare the depletion times and gas fractions of the three CO samples discussed above to the best-fit \citetalias{Tacconi20} scaling relations (their Tables 2 and 3). The properties of these galaxies were measured with similar assumptions to those employed by \citetalias{Tacconi20}. For consistency, we re-scaled all of the \citetalias{Birkin21} $M_{\rm mol}$ measurements to $\alpha_{\rm CO} = 3.6$, which matches our prescription in Section \ref{sec:gas} and that of \citetalias{Aravena19}, as well as the metallicity-based $\alpha_{\rm CO} = 2-5$ used by \citetalias{Tacconi20}.

\subsubsection{Constant Depletion Times}
\label{sec:tdep}

\begin{figure*}[htb]
    \centering
    \includegraphics[width=\linewidth]{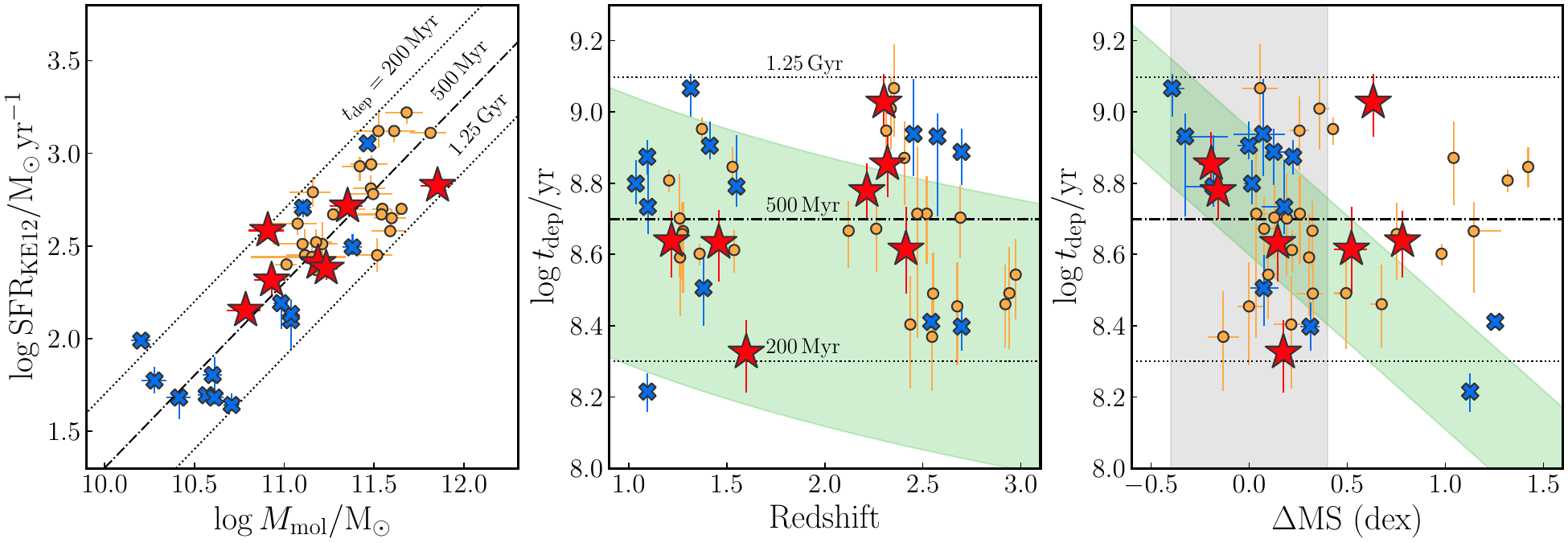}
    \caption{Plots showing the relationship between molecular gas and SFR for $z = 1-3$ CO detections. Symbols and colors are the same as in Figure \ref{fig:main_sequence}. \textit{Left:} SFR vs molecular gas mass. Lines of constant depletion time $t_{\rm dep} \equiv M_{\rm mol} / {\rm SFR} = 0.2$, 0.5, and 1.25\,Gyr are shown. \textit{Middle:} Depletion time vs. redshift. The green shaded region shows the best-fit scaling relation from \citetalias{Tacconi20} for a range $M_\star = 10^{10-12}\,{\rm M}_\odot$ and $\Delta{\rm MS} = -0.2$ to $1.2\,$dex (the 5th-to-95th percentile range in $\Delta$MS for the galaxies shown). \textit{Right:} Depletion time vs. main sequence offset. The grey shaded region is the same as in Figure \ref{fig:main_sequence}. The green region shows the best-fit scaling relation from \citetalias{Tacconi20}, this time for $z = 1-3$ and $M_\star = 10^{10-12}\,{\rm M_\odot}$.}
    \label{fig:tdep}
\end{figure*}

In the left panel of Figure \ref{fig:tdep}, we plot SFR versus $M_{\rm mol}$ for the three CO samples. These properties are tightly correlated, resulting in a small range of depletion times, indicated by the dotted lines of fixed $t_{\rm dep}$ in this panel. We provide $t_{\rm dep}$ for individual galaxies in our sample in Column (9) of Table \ref{tab:properties}; our galaxies have $t_{\rm dep} =210$\,Myr to 1.07\,Gyr. GN-8051 and GN-19876, which do not have CO detections and therefore are not shown in Figure \ref{fig:tdep}, have short depletion times, with $t_{\rm dep} = 260$\,Myr based on [C\,{\sc i}](1--0) in GN-8051, and $t_{\rm dep} \leq 250$\,Myr based on the CO(4--3) upper limit in GN-19876. 

\citetalias{Tacconi20} find that $t_{\rm dep}$ declines at high $z$ and high $\Delta$MS; specifically, they measure $t_{\rm dep,T20} \propto (1+z)^{-0.98} \, (\Delta{\rm MS})^{-0.49}$. We plot $t_{\rm dep}$ against $z$ and $\Delta$MS for the three CO samples in the center and right panels of Figure \ref{fig:tdep}, respectively. The green shaded regions in these panels show the best-fit relation from \citetalias{Tacconi20} for a wide range of galaxy properties, specified in the caption of Figure \ref{fig:tdep}.

In contrast to the \citetalias{Tacconi20} predictions, the depletion times of the CO detections do not appear to evolve with either $z$ or $\Delta$MS. The median (16th-to-84th percentile range) $t_{\rm dep}$ across all three CO samples is 500 (290--850)\,Myr, with the ASPECS-LP sample having slightly longer $t_{\rm dep}$ (median 630\,Myr) compared to our sample (440\,Myr) and that of \citetalias{Birkin21} (460\,Myr). The median depletion times of $z < 2$ galaxies (480\,Myr) and $z > 2$ galaxies (510\,Myr) are consistent with each other, as are those of main-sequence galaxies (520\,Myr) and starbursts (450\,Myr). Pearson correlation tests likewise suggest there is no correlation between $\log t_{\rm dep}/{\rm yr}$ and $\log(1+z)$ (Pearson $r = -0.09$, $p = 0.59$) or $\Delta$MS ($r = -0.21$, $p = 0.17$). In summary, our findings are consistent with a constant depletion time of $\sim$500\,Myr across nearly two orders of magnitude in $M_{\rm mol}$, SFR, and $\Delta$MS.

Because \citetalias{Tacconi20} predict much shorter depletion times in starbursts ($t_{\rm dep} \sim 200$\,Myr for $z \sim 2$ and $\Delta{\rm MS} \sim 1$\,dex), the starbursting galaxies have the largest offsets from the \citetalias{Tacconi20} relation. We measure a mean offset of $\log t_{\rm dep} / t_{\rm dep,T20} = +0.10\pm0.04$\,dex across all samples and $-0.01\pm0.04$\,dex in main-sequence galaxies (errors on the mean are from jackknife resampling). However, we measure a mean offset of $+0.33\pm0.07$\,dex (a factor of 2.1) for just the starbursts, similar to the offsets found in low-mass starbursts by \citet{Tacconi18}. In Section \ref{sec:mu_mol}, we will show that the gas fractions of starbursts are likewise roughly twice as high as the scaling relations predict.

As a final note on the depletion timescale, our measurements of $t_{\rm dep} \sim$500\,Myr are considerably shorter than the $t_{\rm dep} \gtrsim1$\,Gyr previously measured by \citetalias{Aravena19}. This appears to be due to our use of $L_{\rm IR}$-based SFRs, which are systematically lower than the SFRs returned directly from SED-fitting codes (MAGPHYS for ASPECS-LP and \citetalias{Birkin21}, and BAGPIPES for this work). Direct SED-based SFRs result in median depletion times of 1.28, 0.78, 1.24, and 1.13\,Gyr for ASPECS-LP, \citetalias{Birkin21}, our sample, and all samples, respectively. Again, this highlights the need for a self-consistent framework when comparing the properties of galaxies from different works.

\subsubsection{Higher than Expected Gas Fractions in Starbursts}
\label{sec:mu_mol}

\begin{figure*}[htb]
    \centering
    \includegraphics[width=0.85\linewidth]{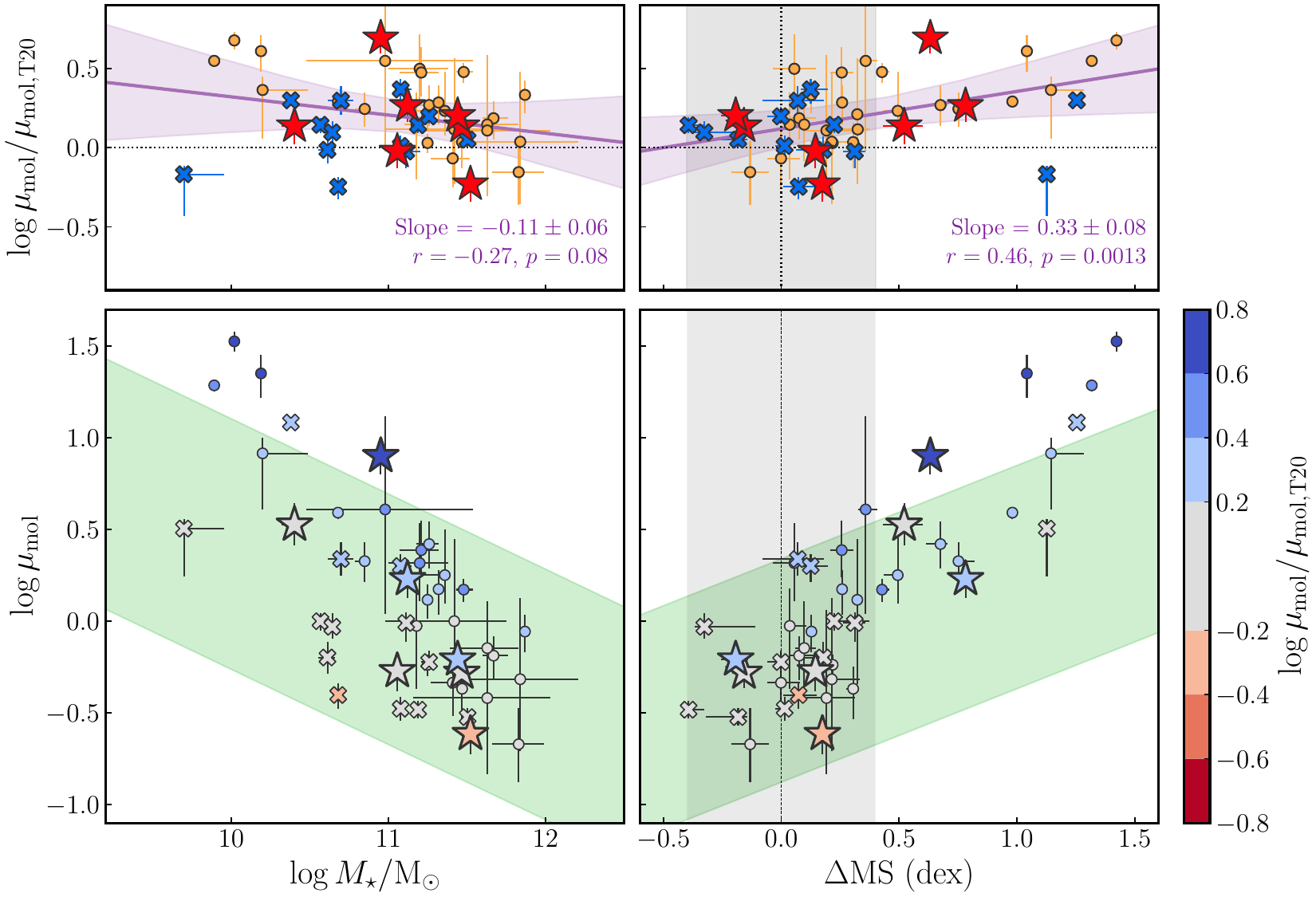}
    \caption{Relationships between molecular gas fraction, $\mu_{\rm mol} = M_{\rm mol} / M_\star$, and galaxy physical properties. Symbols are the same as in Figures \ref{fig:main_sequence} and \ref{fig:tdep}, but colors in the bottom two panels now show offsets of $\mu_{\rm mol}$ from the best-fit \citetalias{Tacconi20} scaling relation---grey points are within the 0.2\,dex scatter of this relation, while blue (red) colors indicate galaxies that are more (less) gas-rich than predicted. The black dashed line and grey shaded region again indicate the \citet{Speagle14} SFMS and a range $\Delta{\rm MS} = -0.4$ to 0.4\,dex, respectively. \textit{Bottom left:} Gas fraction vs. stellar mass. The green shaded region shows the \citetalias{Tacconi20} relation for $z = 1-3$ and $\Delta{\rm MS} = -0.2$ to 1.2\,dex. \textit{Bottom right:} Gas fraction vs. main sequence offset. The green shaded region shows the \citetalias{Tacconi20} relation for $z = 1-3$ and $M_\star = 10^{10-12}\,{\rm M}_\odot$. \textit{Top panels:} Offsets of the measured gas fractions from the predicted values of \citetalias{Tacconi20} vs. stellar mass and main sequence offset. The horizontal dotted lines indicate $\log \mu_{\rm mol} / \mu_{\rm mol,T20} = 0$. The purple lines show the best-fit linear relationships between these variables, with $3\sigma$ errors on this fit shaded purple. The best-fit slope, Pearson $r$ statistic, and $p$ value for the linear correlation are shown in the bottom right corner.}
    \label{fig:mu_mol}
\end{figure*}

In contrast to our small range of depletion times, our faint DSFGs exhibit a wide range of molecular gas fractions, $\mu_{\rm mol} = 0.24-7.95$. We provide $\mu_{\rm mol}$ for individual faint DSFGs in our sample in Column (10) of Table~\ref{tab:properties}. The ASPECS-LP and \citetalias{Birkin21} CO detections likewise span a large range of $\mu_{\rm mol}$. 

We plot $\mu_{\rm mol}$ against $M_\star$ and $\Delta$MS for the three CO samples in the bottom-left and bottom-right panels, respectively, of Figure \ref{fig:mu_mol}. The symbols are the same as in Figures \ref{fig:main_sequence} and \ref{fig:tdep}, but to illustrate better the differences between measured and predicted values of $\mu_{\rm mol}$, we color code the points in the bottom two panels of Figure \ref{fig:mu_mol} according to their offsets from the \citetalias{Tacconi20} relation---galaxies with $\mu_{\rm mol}$ consistent with the relation are colored grey, and galaxies that are more (less) gas-rich than expected are colored blue (red). Similar to Figure \ref{fig:tdep}, the green shaded regions these panels show the best-fit scaling relation of \citetalias{Tacconi20}, $\mu_{\rm mol,T20} \propto M_\star^{-0.41} \, \Delta{\rm MS}^{0.52}$. 

The gas fractions of the CO-detected galaxies do increase at lower $M_\star$ and higher $\Delta$MS, as expected, but do so more rapidly than \citetalias{Tacconi20} predict. In other words, while massive, main-sequence galaxies fit the \citetalias{Tacconi20} relation, starbursting galaxies are more gas-rich than predicted. Specifically, we measure a a mean offset $\log \mu_{\rm mol} / \mu_{\rm mol,T20} = +0.13 \pm 0.03$\,dex in the main-sequence galaxies, and $\log \mu_{\rm mol} / \mu_{\rm mol,T20} = +0.35\pm0.06$\,dex (a factor of 2.2) in the starbursts. The latter value is consistent with our finding in Section \ref{sec:tdep} that starbursts have $t_{\rm dep}$ 2.1 times longer than predicted.

Figure \ref{fig:mu_mol} also shows that galaxies with lower stellar masses are more gas-rich than \citetalias{Tacconi20} predict; however, because of the biased relationship between $M_\star$ and $\Delta$MS (see Section \ref{sec:main_sequence}), it is difficult to determine whether this trend is simply driven by the fact that the low-mass galaxies are also starbursts. We attempted to test this by performing simple linear regressions of $\log \mu_{\rm mol} / \mu_{\rm mol,T20}$, which we show with the purple lines in the top panels of Figure \ref{fig:mu_mol}. We find a significant correlation (Pearson $r = 0.46$, $p = 0.0013$) between $\log \mu_{\rm mol}/\mu_{\rm mol,T20}$ and $\Delta$MS, but only a weak anti-correlation ($r = -0.27$, $p = 0.08$) between $\log \mu_{\rm mol}/\mu_{\rm mol,T20}$ and $\log M_\star/{\rm M}_\odot$. The difference is mainly due to the ASPECS-LP galaxies, which comprise the majority of $M_\star < 10^{11}\,{\rm M}_\odot$ galaxies among the three CO samples, but only two out of 13 starbursts, and are largely consistent with the \citetalias{Tacconi20} relations. Further observations of galaxies with $M_\star \sim 10^{10}\,{\rm M_\odot}$ both on and above the SFMS will be necessary to accurately measure the evolution of $\mu_{\rm mol}$ with both $M_\star$ and $\Delta$MS.

\subsubsection{Evidence for Evolving $\alpha_{\rm CO}?$}

The factor of $\approx$2 higher $t_{\rm dep}$ and $\mu_{\rm mol}$ in low-mass and starbursting galaxies than predicted by the \citetalias{Tacconi20} scaling relations may indicate that these relations do not hold for low-mass starbursts. However, this may also indicate that $M_{\rm mol}$ is systematically overestimated in this population, i.e., by not accounting for the stellar mass- or sSFR-dependence of $\alpha_{\rm CO}$ and/or CO excitation. Although a full analysis of these effects is outside the scope of this work, we briefly comment on these possible factors below.

First, $\alpha_{\rm CO}$ is known to vary widely depending on ISM conditions that can be correlated with both $M_\star$ and $\Delta$MS \citep[see, e.g., the review by][]{Bolatto13}. The observed $t_{\rm dep}$ and $\mu_{\rm mol}$ offsets could be accounted for by a lower CO-to-H$_2$ conversion factor, $\alpha_{\rm CO} \sim 1-2$, in the starburst galaxies \citep[e.g.,][]{Magnelli12, Bothwell13}. Notably, applying a metallicity correction to $\alpha_{\rm CO}$ \citep[e.g.,][]{Genzel12, Genzel15, Tacconi18} by assuming a mass-metallicity relation would \textit{worsen} the deviations, since this would increase $\alpha_{\rm CO}$ in the low-mass galaxies and decrease $\alpha_{\rm CO}$ in the more massive galaxies that already fit the \citetalias{Tacconi20} relations. A more nuanced approach to $\alpha_{\rm CO}$ assumptions is necessary.

The CO in these galaxies could also be more excited, such that the luminosities of the detected $J_{\rm up} \geq 2$ CO lines correspond to lower $L'_{\rm CO(1-0)}$ values. \citet{Riechers20} found that assumptions about CO excitation alone could bias molecular gas mass estimates systematically by up to a factor of 2. \citet{FriasCastillo23} did not find evidence for correlation between CO excitation and galaxy properties including SFR surface density, but their sample was limited to very bright DSFGs with $S_{\rm 850 \mu m} > 10$\,mJy at $z = 2-5$.

Finally, the selection biases in the left panel of Figure~\ref{fig:main_sequence} make disentangling the joint effect of $M_\star$ and $\Delta$MS on $\alpha_{\rm CO}$ and the CO SLED difficult. Further CO observations in DSFGs both on and above the SFMS, as well as a careful approach to measuring physical properties consistently, are needed.

\subsection{Red Selection for CO in Low-$M_\star$ Main Sequence DSFGs}
\label{sec:future}

In Section \ref{sec:main_sequence}, we showed that sensitivity limits of submm/mm continuum surveys prevent the selection of DSFGs on the SFMS at $\lesssim 10^{11}\,M_\odot$ for follow-up CO observations. This is problematic, given the expense of blind surveys like ASPECS-LP. Intentionally targeting main sequence galaxies \citep[e.g.,][]{Tacconi13, Genzel15} at low mass from submm parent samples will require substantially deeper continuum observations. 

The flux limits (and thus SFR limits) of single-dish submm/mm surveys will be vastly improved by sensitive cameras on larger telescopes than the 15-m JCMT, such as the NIKA2 camera on the IRAM 30-m Telescope \citep[e.g.,][]{Bing23}, or the TolTEC camera on the Large Millimeter Telescope (LMT; \citealp{Wilson20}). The TolTEC survey of the UDS, for example, has an anticipated 4$\sigma$ detection threshold of $L_{\rm IR} \sim 10^{11}\,L_\odot$ \citep{NavaMoreno24}, or ${\rm SFR} \sim 14\,M_\odot\,{\rm yr}^{-1}$ using Equation \ref{eqn:ir_sfr}---over an order of magnitude lower than that of SCUBA-2 at 850\,$\mu$m. We plot this limit for $z = 2.5$ as a magenta dot-dashed line in Figure \ref{fig:main_sequence}, which shows that instruments like TolTEC will enable the targeted follow-up of large samples of low-mass ($\log M_\star / M_\odot \gtrsim 9-9.5$) main sequence DSFGs around cosmic noon for the first time. Importantly, while some mosaicked interferometric continuum surveys can achieve similar sensitivities \citep[e.g.,][]{GonzalezLopez20}, the large areas covered by single-dish surveys are necessary to minimize bias due to cosmic variance.

Intentionally targeting low-mass DSFGs both on and above the SFMS will require accurate counterpart matching between the submm/mm sources and their OIR counterparts. At a minimum, any method used for counterpart selection must have better spatial resolution than the submm/mm continuum imaging, and it must select at least as many galaxies as the number of submm/mm sources. These requirements make most existing methods of counterpart selection unusable for the next generation of submm/mm surveys. 

For the TolTEC UDS survey, the expected surface density of submm/mm detections above their $L_{\rm IR} \sim 10^{11}\,{\rm L}_\odot$ threshold is $\sim$10\,arcmin$^{-2}$, and the spatial resolution of TolTEC at 1.1\,mm is $\sim$5$''$ \citep{NavaMoreno24}. Radio counterparts (3.1\,arcmin$^{-2}$ for $S_{\rm 1.4GHz} > 11\,{\rm \mu Jy}$; \citealp{Owen18}) and the F160W--IRAC2 red selection used in this work ($\sim$5\,arcmin$^{-2}$) do not provide enough galaxies to account for the stellar counterparts of these sources. MIR point sources are more numerous (16 arcmin$^{-2}$; \citealp{Elbaz11}), but the spatial resolution of Spitzer/MIPS at 24\,$\mu$m is not an improvement over that of TolTEC itself. The best option for identifying the stellar counterparts of the faintest DSFGs may instead be NIRCam-based red selection \citep{Barger23}, since NIRCam imaging in deep legacy fields, such as the GOODS-N, is already $\sim$3\,mag deeper at 1.5\,$\mu$m and $\sim$5.5\,mag deeper at 4.5\,$\mu$m than the equivalent HST and \textit{Spitzer} filters \citep{Eisenstein23}.

\section{Summary}
\label{sec:summary}

We presented our NOEMA survey of CO(3--2) and CO(4--3) in eight DSFGs near the SCUBA-2 confusion limit ($S_{\rm 850 \mu m} = 1.4-3.5$\,mJy) in the CANDELS GOODS-N field. We selected the HST counterparts to these galaxies using a HST/Spitzer NIR flux-color cut, and we limited the sample to galaxies with existing spectroscopic redshifts from the literature. This allowed us to target the CO (or [C\,{\sc i}]) line in each galaxy with a single NOEMA frequency tuning. Seven out of our eight targets were detected in CO at a $>5\sigma$ significance at the expected frequency based on their OIR redshifts. We additionally detected the 2\,mm continuum in two of our targets, as well as in two serendipitous off-axis DSFGs, one of which was also detected in [C\,{\sc i}](1--0).

We derived molecular gas masses from each of the emission lines, performed UV-to-mm SED fits to measure $L_{\rm IR}$ and $M_\star$ for our sample, and calculated SFRs assuming the SFR--$L_{\rm IR}$ conversion from \citetalias{Kennicutt12}. Finally, we compared the results for our faint DSFG sample, along with those for two other faint, CO-detected DSFG samples at $z = 1-3$ from the literature, to the SFMS of \citet{Speagle14} and the molecular gas scaling relations from \citetalias{Tacconi20}. Our conclusions from this work are as follows:

\begin{enumerate}
    \item Our NIR color selection of $f_{\rm 4.5\mu m}/f_{\rm F160W} > 3.5$ and $f_{4.5\mu m} > 1\,\mu$Jy correctly identified the OIR counterparts for seven  of our eight NOEMA targets, which is consistent with the $\sim$90\% accuracy of similar selections in the literature.
    \item Our intentional selection of the faintest SCUBA-2 sources in the GOODS-N allowed us to detect CO in galaxies with lower SFRs than those in similar follow-up surveys with NOEMA and ALMA.
    \item The DSFGs in this work, both from our survey and from the two literature samples, are all classified as either main sequence galaxies or starburst galaxies based on the assumptions used in this work. We caution that comparisons of DSFGs across surveys, especially their offsets from the SFMS, requires careful consideration of the SFMS and SFR prescriptions used.
    \item Due to the direct relationship between submm flux and SFR, follow-up CO surveys derived from submm flux-limited parent samples display a selection bias, such that low-mass ($M_\star \lesssim 10^{11}\,{\rm M}_\odot$) galaxies on the main sequence are absent from these samples.
    \item Galaxies on the SFMS at $z = 1-3$ have depletion timescales and molecular gas fractions consistent with molecular gas scaling relations from \citet{Tacconi20}, but low-mass, starbursting DSFGs have depletion timescales and gas fractions a median of two times higher than predicted, which may be evidence for lower $\alpha_{\rm CO}$ and/or higher CO excitation in these galaxies.
    \item Disentangling the $M_\star$- and $\Delta$MS-dependences of $\alpha_{\rm CO}$ and the CO SLED will require further observations of CO in DSFGs on and above the SFMS, which may be identified via a NIRCam-based selection.
\end{enumerate}

\begin{acknowledgements}

This work is based on observations carried out under project numbers W23CH and S24BR with the IRAM NOEMA Interferometer. IRAM is supported by INSU/CNRS (France), MPG (Germany) and IGN (Spain). M.J.N.R. thanks our IRAM contacts, Kirsty May Butler and Romane Le Gal, for valuable assistance in the reduction of these NOEMA data.

The James Clerk Maxwell Telescope is operated by the East Asian Observatory on behalf of The National Astronomical Observatory of Japan; Academia Sinica Institute of Astronomy and Astrophysics; the Korea Astronomy and Space Science Institute; the National Astronomical Research Institute of Thailand; Center for Astronomical Mega-Science (as well as the National Key R\&D Program of China with No. 2017YFA0402700). Additional funding support is provided by the Science and Technology Facilities Council of the United Kingdom and participating universities and organizations in the United Kingdom and Canada.

\end{acknowledgements}

%

\facilities{JCMT (SCUBA-2), NOEMA, Keck-I (LRIS, MOSFIRE), HST (ACS, WFC3), SMA, VLA, Subaru (SuprimeCam), CFHT (WIRCam), Herschel (SPIRE), Spitzer (IRAC).}


\software{astropy \citep{astropy13, astropy18, astropy22}; \textsc{gildas} \citep{GILDAS}; \texttt{numpy} \citep{Harris20}; \texttt{scipy} \citep{Virtanen20}} 

{}

\end{document}